\documentclass[12pt]{article}
\pdfoutput=1
\usepackage{jheppub}
\usepackage{amsmath}
\usepackage{graphicx}
\usepackage[font=small]{caption,subcaption}
\usepackage{cancel}
\usepackage{mathrsfs}
\usepackage{amssymb}
\usepackage{epstopdf}
\usepackage{dsfont}
\usepackage{float}
\usepackage{amsthm}
\usepackage{braket}
\usepackage{lmodern}
\usepackage{mathtools}
\newcommand\numberthis{\addtocounter{equation}{1}\tag{\theequation}} 
\usepackage{empheq} 

\usepackage{slashed}
\usepackage{bbm}
\usepackage{subcaption}	
\usepackage{braket}
\usepackage{bm}
\usepackage{soul}

\newcommand{\DD}{\mathrm{d}}
\newcommand{\defined}{\coloneqq}		
\newcommand{\e}{\epsilon}

\renewcommand{\d}{\partial}
\newcommand{\A}{\mathcal{A}}

\renewcommand{\.}{\cdot}

\newcommand{\I}{\mathbbm{1}}

\renewcommand{\O}{\mathcal{O}}

\usepackage{xfrac}						
\usepackage{breqn}
\newcommand{\at}[1]{\bigg\vert_{#1}}

\graphicspath{{./figures/}}

\addtocontents{toc}{\protect\enlargethispage*{3\baselineskip}} 

\newpage

\title{Zwanziger's Pairwise Little Group on the Celestial Sphere}

\author{Luke Lippstreu}
\affiliation{Department of Physics, Brown University, Box 1843, Providence, Rhode Island 02912, USA}

\emailAdd{luke\_lippstreu@brown.edu}

\abstract{
We generalize Zwanziger's pairwise little group to include a boost subgroup. We do so by working in the celestial sphere representation of scattering amplitudes. We propose that due to late time soft photon and graviton exchanges, matter particles in the asymptotic states in massless QED and gravity transform under the Poincar\'e group with an additional pair of \textit{pairwise} celestial representations for each pair of matter particles. We demonstrate that the massless abelian and gravitational exponentiation theorems are consistent with the proposed pairwise Poincar\'e transformation properties. For massless QED we demonstrate that our results are consistent with the effects of the Faddeev-Kulish dressing and the abelian exponentiation theorem for celestial amplitudes found in \href{https://arxiv.org/abs/2012.04208}{2012.04208}.  We discuss electric and magnetic charges simultaneously as it is especially natural to do so in this formalism. }

\begin{document}
\maketitle
\section{Introduction}
The traditional formulation of the $S$-matrix assumes that the asymptotic states transform as a tensor product of Wigner's one-particle irreducible representations of the Poincar\'e group \cite{Wheeler:1937zz,Wigner,Weinberg:1995mt}.  However, when interactions are mediated by massless particles, matter particles continue to interact at asymptotic times via the exchange of soft massless modes, and hence never completely decouple. As such, the asymptotic states do not transform as one-particle irreducible representations of the Poincar\'e group. This particular inconsistency is the root cause of infrared (IR) divergences in the S-matrix.  Thus the question naturally arises:\\
\newline
\textit{To which unitary representations of the Poincar\'e group do the asymptotic states in abelian gauge theories and gravity correspond?}\\
\newline
In this paper we propose a partial\footnote{We expect that a complete description of the Poincar\'e transformation properties of the asymptotic states will require one to also specify the in/out vacuum states, similar to those given in \cite{A4He:2014cra,Kapec:2017tkm}. Herein we limit our discussion to the non-vacuum sector.} answer to this question for massless matter particles.  Using the insights made in the recent analyses \cite{Csaki:2020inw,Csaki:2020uun}, which fully clarified the representation theory of Zwanziger's pairwise little group \cite{Zwanziger:1972sx}, our work generalizes Zwanziger's pairwise little group to include a boost subgroup. Zwanziger's original pairwise little group accurately captures the novel rotation properties of an electric-magnetic charge pair, and we demonstrate that our generalization suggests novel boost transformation properties for matter particles in abelian gauge theories and gravity. This generalization is most readily achieved by working in the celestial sphere representation of scattering amplitudes \cite{Strominger:2013lka,Pasterski:2016qvg,Pasterski:2017kqt,Banerjee:2018gce}. We discuss the technically more challenging case of massive matter particles in the four-momentum basis in \cite{L.Lippstreu}.\par
We raise the above question specifically for abelian gauge theories and gravity not only due to the physical relevance of these theories, but also due to their especially simple infrared structure. This simplicity is exhibited in the abelian and gravitational exponentiation theorems \cite{Yennie,Weinberg:1965nx}, which state that resumming the leading order infrared divergent contributions to scattering amplitudes due to soft virtual photon and graviton exchanges between external legs of Feynman diagrams results in the amplitudes being multiplied by universal exponential prefactors. Moreover, the exponents of these exponentials take the form of a pairwise sum over the matter particles in the in/out states. These theorems seem to be suggesting a simple underlying covariance to the asymptotic states, in which the transformation law is at most pairwise for the particles involved.  In this paper we propose such a pairwise Poincar\'e transformation law and demonstrate that the soft factors in the gravitational and abelian exponentiation theorems are the amplitudes which exhibit the proposed covariance. \par
The simplicity of IR divergences in abelian gauge theories and gravity should be contrasted with the corresponding situation in non-abelian gauge theories where the IR divergences do still factorize off from the hard scattering in a universal form \cite{Gatheral:1983cz,Frenkel:1984pz,Collins:1988ig,Feige:2014wja,Gardi:2013ita}, but the resultant structure is not as compact as in the abelian and gravitational cases, and the universal factors no longer take a pairwise form, but instead involve so-called webs of multiple particles.   \par
To test the utility of our generalization of Zwanziger's pairwise little group, we compare our results to the recent study \cite{Arkani-from uv to IR} of infrared divergences on the celestial sphere. Therein the abelian exponentiation theorem was translated into an equivalent statement for celestial amplitudes of massless charged particles. The celestial states were then dressed with an appropriate choice of Faddeev-Kulish dressing \cite{Kulish:1970ut}.  It was concluded that the net effect of these two infrared contributions was simply to shift the conformal dimensions of the charged particle states in the amplitude by an amount $\frac{e^2}{4\pi^2}\ln\Lambda_{IR}$, where $e$ denotes the state's electric charge and $\Lambda_{IR}$ refers to the infrared cutoff. In this paper we demonstrate that Zwanziger's pairwise little group, when generalized to the celestial sphere, provides an alternative rationale for this result\footnote{Recently in \cite{gluonIRshift} the hard/soft factorization properties of gluon amplitudes were converted into corresponding statements for celestial amplitudes. Again it was found that the soft divergences resulted in a shift of the conformal dimension of the celestial operators by a divergent amount proportional to the cusp anomalous dimension. This indicates that aspects of our analysis may also be relevant for non-abelian theories.}. \par
Our understanding of the infrared sector of abelian gauge theories has a long history, beginning in 1937 with the work of Bloch and Nordsieck \cite{Bloch:1937pw}. Therein it was realized that infrared divergences could be made to cancel on the level of inclusive cross sections by summing over experimentally indistinguishable processes in which soft photons may have been radiated away undetected.  This procedure was completely systematized in the work of Yennie et al. \cite{Yennie}.
 However, the Bloch-Nordsieck approach is formulated at the level of inclusive cross sections, whereas it is desirable to have a procedure that works at the level of amplitudes so that properties such as unitarity can be better studied. This motivated Chung \cite{Chung:1965zza} and Kibble \cite{Kibble1,Kibble:1969ep,Kibble:1969ip,Kibble:1969kd} to develop their so-called coherent state approach where the in/out states are modified to consist of charged particles surrounded by an indefinite number of photons. The works of Chung and Kibble however did not take into account the so-called Coulomb phase divergence. This divergence was first encountered in non-relativistic Coulomb scattering by Dalitz \cite{Dalitz}. The modern treatment of this divergence was systematized by Dollard in \cite{Dollard}, who demonstrated that one can define an IR finite Coulomb scattering matrix by using distorted plane waves as the in/out states. This spur of activity during the 1960's culminated in the work of Faddeev and Kulish (FK) \cite{Kulish:1970ut} who unified the works of Chung, Kibble and Dollard. In the FK construction one accounts for the fact that the asymptotic evolution of charged states is not free evolution, but instead involves a continual radiation and exchange of soft photons. FK then define the infrared finite $S$-matrix to be the part of the Dyson $S$-matrix with the asymptotic evolution factored out. The extension of the FK formalism to the case of massless charged particles, as well as for non-abelian gauge theories, was recently developed in \cite{Hofi}. As far as we are aware, the Lorentz covariance properties of the FK states are not yet known, except in the specific case of a celestial state dressing, where a conformal primary condition picks out a particularly simple choice of FK dressing \cite{Arkani-from uv to IR}.\par
 The necessity for a type of soft photon dressing has also been motivated by considerations of gauge invariance. In \cite{Dirac:1955uv} Dirac sought a gauge invariant formulation of quantum electrodynamics. This naturally lead them to dress the charged particle states with a surrounding Coulomb field of photons.  The relation between gauge invariance and the dressing of the asymptotic states has been further pursued in \cite{Bagan:1999jf,Bagan:2001uq,Hirai:2019gio}. \par 
  An alternate viewpoint on the origin of IR divergences has recently emerged in studies of asymptotic symmetries \cite{A1Strominger:2013jfa,A2He:2014laa,A3Kapec:2014opa,A4He:2014cra}. In \cite{Kapec:2017tkm} it was demonstrated that one could view IR divergences in QED as arising from the traditional $S$-matrix failing to account for the vacuum transitions between scattering events, which must necessarily occur in order for the asymptotic symmetries to be respected. \par
All of these approaches to infrared divergences can be viewed as emphasizing various different aspects of the same underlying cause; the asymptotic states do not transform as free particle representations. This leads us to suspect that further insights into the IR structure of abelian gauge theories and gravity can be gained from a more complete understanding of the Poincar\'e transformation properties of the asymptotic states. \par
As the asymptotic states in abelian gauge theories and gravity never completely decouple, one requires a formulation that goes beyond a simple tensor product of one-particle irreducible representations. We believe that Zwanziger made a fundamental advance in this regard in their severely underappreciated work on monopoles \cite{Zwanziger:1972sx}. In the case of an electric-magnetic charge pair, there remains a quantized amount of angular momentum stored in the electromagnetic field, regardless of the pair's separation distance. Consequently the rotation properties of the asymptotic states are not those of free particle states. Zwanziger generalized Wigner's representations of the Poincar\'e group in a manner that correctly accounted for the asymptotic dynamics of the electric-magnetic charge pair. Zwanziger's fundamental insight was the observation that any pair of momenta always has a $U(1)$ little group in common. For example, one can always Lorentz boost to a frame where the pair are traveling along the $z$-axis. In this frame both momenta are unchanged by a rotation about the $z$-axis.  We will refer to the group generated by the algebra $\set{J_z}$, where $J_z$ denotes the generator of rotations about the $z$-axis, as the particles' \textit{pairwise} little group.  Utilizing this additional $U(1)$ pairwise little group Zwanziger was able to correctly account for the additional angular momentum stored in the electromagnetic field. With the asymptotic transformation properties of the electric-magnetic pair understood, Zwanziger was able to calculate some of the scattering amplitudes of dyons with relative ease, a problem which is otherwise fraught with complications \cite{Zwanziger:1970hk,monopoleproCsaki:2010rv,monopoleprobBruemmer:2009ky,monopoleprobColwell:2015wna,monopoleprobGamberg:1999hq,monopoleprobHook:2017vyc,monopoleprobSanchez:2011mf,monopoleprobTerning:2018lsv,monopoleprobTerning:2018udc,monopoleprobWeinberg:1965rz}.  We review Zwanziger's construction in more detail in Section \ref{Zwanzigersummarysect}. More complete details can be found in the original work {\cite{Zwanziger:1972sx}}, as well as the recent studies \mbox{\cite{Csaki:2020inw,Csaki:2020uun}} which resurfaced Zwanziger's work by recasting the construction in terms of the massive spinor-helicity formalism of {\cite{Arkani-Hamed:2017jhn}} by introducing a new pairwise spinor-helicity variable, thereby streamlining the study of the three point and partial wave amplitudes of electric-magnetic particle scattering. One of the reasons why Zwanziger's work has remained dormant for so long may be due to the equivocal representation theory of the additional $U(1)$ phase. The recent works \cite{Csaki:2020inw,Csaki:2020uun} clarified the crucial point that the additional phase should be thought of as arising from the Poincar\'e group acting an \textit{additional} pairwise state, over and above the tensor product of two particle states that Zwanziger wrote down. The useful nomenclature of a pairwise state and a pairwise helicity, used throughout in this paper, also originates from these recent works.   \par
In the case of a electric-magnetic charge pair there remains an angular momentum between the two at asymptotic times, beyond that of the usual orbital and spin angular momentum. We argue in Section \ref{sectboostcharge} that for an electric charge pair, an analogous situation holds, where there remains a \textit{boost} charge between the two at asymptotic times, beyond that of the free theory's boost charge. This motivates us to search for a construction where a \textit{pairwise} little group that contains boosts can be found. Such a pairwise little group can be found if we rewrite the particle states in terms of the so-called celestial sphere basis \cite{Pasterski:2016qvg,Pasterski:2017kqt}. This basis has recently come to the fore in studies of asymptotic symmetries \cite{A1Strominger:2013jfa,A2He:2014laa,A3Kapec:2014opa,A4He:2014cra} where it appears to be the appropriate basis for elucidating the asymptotic symmetry groups of gauge theories and gravity \cite{Strominger:2013lka,c1Pasterski:2017ylz,c2Lam:2017ofc,c3Schreiber:2017jsr,c4Stieberger:2018edy,c5Stieberger:2018onx,c6Fan:2019emx,c7Pate:2019mfs,c8Puhm:2019zbl,c9Nandan:2019jas,c10Adamo:2019ipt,c11Guevara:2019ypd,c12Pate:2019lpp,c13Law:2019glh,c14Donnay:2020guq,c15Albayrak:2020saa,c16Casali:2020vuy,c165Casali:2020uvr,c18Kapec:2014opa,c19He:2014cra,c20Strominger:2014pwa,c21He:2015zea,c22Pasterski:2015zua,c23Muck:2020wtx,c24Narayanan:2020amh,c25Banerjee:2020zlg,c26Gonzalez:2020tpi,c27Cachazo:2014dia,c30Choi:2017ylo,c31Donnay:2018neh,c32StromLec}. There are two often-used routes for arriving at the celestial sphere basis, and we review both in Section \ref{sectsummaryofcelestial}. The first is by Mellin transforming on the energy variables of ordinary one-particle states. This route highlights that the celestial states are indeed just a change of basis from the momentum basis, and thus knowledge of celestial amplitudes constitutes knowledge of ordinary Minkowski amplitudes. The other route is by inducing from the so-called lower triangular group of the Lorentz group \cite{Banerjee:2018gce}. In this approach the little group of a massless particle traveling along the $z$-axis is generated by the set of algebra elements $\set{J_z,K_z,J_y-K_x,J_x+K_y}$, which is the ordinary little group of a massless particle appended by boosts along the $z$-axis $K_z$. Given any pair of celestial states we can always Lorentz transform to a frame where one of the particles is pointing along the positive $z$-axis, and the other along the negative $z$-axis. In this frame, the intersection of the little groups of the pair is the group generated by $\set{J_z,K_z}$; that is both rotations about, and boosts along the $z$-axis leave the $\pm \hat{z}$ directions invariant. Thus in the context of celestial amplitudes, Zwanziger's pairwise little group is enhanced to include boosts.  We will demonstrate in Sections \ref{sectcompare} and \ref{sect,maingravity} that this additional pairwise, boost-like, little group reproduces some of the known infrared effects that occur in abelian gauge theories and gravity.  \par
The paper is organized as follows. In Section \ref{sectboostcharge} we argue that there remains a boost charge between pairs of charged particles at asymptotic times, implying that the asymptotic states deviate from a free particle transformation law. This motivates us to search for a pairwise, boost-like, enhancement to Wigner's one-particle irreducible representations. Section \ref{sectsummaryofbackground} summarizes all relevant background material: Section \ref{Zwanzigersummarysect} summarizes Zwanziger's description of monopoles and Section \ref{sectsummaryofcelestial} summarizes the representation theory of celestial states. In Section \ref{sectcelestialzwanziger} we generalize Zwanziger's pairwise little group to the context of celestial states. In Section \ref{sectamplitudes} we construct amplitudes consistent with the new transformation law. In Section \ref{sectcompare} we compare to known infrared regulated amplitudes on the celestial sphere \cite{Arkani-from uv to IR}. We discuss the pairwise representation in the four momentum basis in Section \ref{sect,maingravity}. We delay the discussion of gravity until Section \ref{sect,gravity} as the effects of soft graviton exchanges are more simply discussed in the four-momentum basis rather than in the celestial sphere basis. In Section \ref{sect,gravity} we demonstrate that the gravitational and abelian exponentiation theorems are consistent with the proposed pairwise Poincar\'e covariance. In Section \ref{sect,vertex} we indicate that our additional pairwise celestial states are created by the vertex operators of the Goldstone bosons of the spontaneously broken asymptotic symmetries of abelian gauge theories and gravity.
\section{Motivation: The asymptotic boost operator}\label{sectboostcharge}
The interaction between an electric-magnetic charged pair leads to a quantized amount of angular momentum to be stored in the electromagnetic field at asymptotic times. This suggests that the rotation properties of the pair is not that of free particles'. This observation was the starting point for Zwanziger to develop their representation theory of electric-magnetic charge pairs \cite{Zwanziger:1972sx}. In this section we perform a similar calculation for an electric-electric charge pair and demonstrate that there is an additional boost charge at asymptotic times. We then argue that it is the boost operator in particular which causes the asymptotic states in QED to deviate from their free particle representations. This will guide us in future sections to construct an enhancement of Wigner's one particle irreducible representations which is particularly sensitive to boosts. \par
In the presence of interactions we modify the Hamiltonian operator
\begin{equation}
H=H_0+V,
\end{equation}
where $H_0$ is the free theory Hamiltonian and $V$ is the interaction potential. Then in order to preserve the Poincar\'e algebra it is necessary to modify some of the other generators. For example, in the case of magnetic monopoles we would modify the rotation generators, as well as some of the other generators. In Dirac's instant form of dynamics one modifies the boost generators \cite{Dirac:1949cp}
\begin{equation}
\vec{K}=\vec{K}_0+\vec{W},
\end{equation}
where $\vec{K}_0$ is the free particle boost generator. Let the interaction potential be written as a local density
\begin{equation}
V(t)=\int\DD^3 x\,\,\,\mathcal{I}(\vec{x},t).
\end{equation}
Then a choice of $\vec{W}$ that will preserve the Poincar\'e algebra is \cite{Weinberg:1995mt,Dirac:1949cp}
\begin{equation}
\vec{W}(t)=-\int\DD^3x\,\,\vec{x}\,\,\mathcal{I}(\vec{x},t).
\end{equation}
 with the time dependence arising from using the interaction picture.  In order to claim that the states at $t\rightarrow \pm\infty$ transform as free particle representations one requires that $V(t),\vec{W}(t)$ effectively vanish at asymptotic times. In this context, we can say that an interaction picture operator effectively vanishes at asymptotic times if its late time matrix elements between smooth superpositions of $H_0$ eigenstates $\ket{\psi_{\alpha}},\ket{\psi_{\beta}}$ exhibits no poles of the form $\frac{1}{E_{\alpha}-E_{\beta}}$, where $E_i$ denotes the energy eigenvalue \cite{Weinberg:1995mt}. In theories where massless particles can be exchanged this condition is generically not met and consequently the asymptotic states do not transform as free particle representations. Let us now examine the $\vec{W}$ operator in quantum electrodynamics (QED)
\begin{equation}
\vec{W}(t)=-e\int\DD^3x\,\,\vec{x}\,\,\bar{\psi}\gamma^{\mu}\psi A_{\mu}.\label{classicalboostoperator}
\end{equation}
To get better intuition for the late time behaviour of this operator we evaluate it's classical expectation value in the situation where we have two outgoing charged particles $i$ and $j$ represented by the currents
\begin{equation}
J^{\mu}_{i}(\vec{x},t)=e_iu^{\mu}_{i}\int_0^{\infty}\DD\tau\,\,e^{-\e \tau}\,\,\delta^{4}\big(x^{\mu}-u^{\mu}_{i}\tau\big),
\end{equation}
where $\e\rightarrow 0$ is a regulator, $u_i^{\mu}$ is the velocity $u_{i}^{\mu}\defined \frac{p_i^{\mu}}{m_i}$, and the integration bounds are chosen as appropriate for an out state. The gauge field $A^{\mu}_{j}$ sourced by particle $j$ is\footnote{We work in mostly minus signature $\eta=(+,-,-,-)$ throughout.} (working in Lorentz gauge $\d_{\mu}A^{\mu}=0$ and using retarded boundary conditions)
\begin{equation}
A^{\mu}_j(\vec{x},t)=\frac{e_j}{4\pi}\frac{u_j^{\mu}}{\sqrt{(u_j\.x)^2-x^2}}.\label{coulomb}
\end{equation}
We then compute the value of $\vec{W}_{ij}(t)$ in (\ref{classicalboostoperator}) by replacing $\bar{\psi}\gamma^{\mu}\psi\rightarrow J^{\mu}_i$ and $A^{\mu}\rightarrow A^{\mu}_j$ and find that
\begin{align}
\vec{W}_{ij}(t)&=-\int\DD^3 x \,\,\vec{x}\,\, J^{\mu}_iA_{\mu,j}\\
&=-\frac{\vec{p}_j}{E_j}\frac{e_ie_j}{4\pi\beta_{ij}}\label{asymboost}
\end{align}
where
\begin{equation}
\beta_{ij}\defined \sqrt{1-\frac{m_i^2m_j^2}{(p_i\.p_j)^2}}.
\end{equation}
The most interesting aspect of the calculation (\ref{asymboost}) is that the $t$ dependent factors that occurred at intermediate stages canceled amongst one another. We can then trivially take the late time limit and observe that $\lim\limits_{t\rightarrow \infty}W_{ij}(t)$ is non-zero\footnote{The electromagnetic field's angular momentum tensor may provide an additional contribution to the boost operator, however the relevant integrals are ambiguous and depend on the order of integration \cite{Zwanziger:1972sx}.}\textsuperscript{,}\footnote{The radiation field produced during a scattering event provides an additional contribution to $\vec{W}(t)$ which is also non-zero in the late time limit.}\textsuperscript{,}\footnote{The analysis in this section is incomplete as we have neglected ``surface charge" contributions to the boost operator \cite{Campiglia:2017mua,Henneaux:2018gfi}.  We expect that a more complete analysis will yield the same motivation which we seek to convey here, namely that it is the boost operator in particular which causes the asymptotic states to deviate from free particle representations in abelian gauge theories. }. This suggests that the asymptotic boost operator does not coincide with the free field theory boost operator, and consequently that the asymptotic states do not transform as free particle representations under boosts. This would not have been the case if the gauge field had a mass $m$, in which case our result (\ref{asymboost}) would have been damped by a factor of $e^{-m t}$. Let us also note the pairwise nature of the boost charge (\ref{asymboost}). This suggests that the asymptotically interacting representations may exhibit a pairwise transformation law. \par 
Let us examine the massless limit of our result (\ref{asymboost}). Taking $\beta\rightarrow 1$ we then find that each massless charged particle $i$ carries a boost charge
\begin{equation}
\vec{W}_{ij}=-\frac{e_ie_j}{4\pi}\hat{n}_i,\label{masslessboostcharge}
\end{equation} 
which is a function only of the direction of the massless particle $\hat{n}_i=\frac{\vec{p}_i}{E_i}$ and not it's energy. As a look ahead, let us note that celestial states $\ket{z;\triangle,\sigma}$ are eigenstates of boosts in their direction of motion, with an eigenvalue proportional to their conformal dimension $\triangle$. Indeed, what we find in this paper is that Zwanziger's procedure, when applied to celestial states, suggests that the net effect of late time interactions for massless charged particles is to add to each pair of massless charged particles a pair of celestial states with conformal dimension $\triangle_{ij}=\frac{e_ie_j}{4\pi^2}\log\Lambda_{IR}$.  
\section{Review of background material}\label{sectsummaryofbackground}
Our work generalizes Zwanziger's description of magnetic monopoles to also include purely electric interactions. This generalization is most easily achieved in the celestial sphere representation of scattering amplitudes. We briefly review Zwanziger's construction in Section \ref{Zwanzigersummarysect}, and then review the representation theory of celestial states in Section \ref{sectsummaryofcelestial}. 
\subsection{Zwanziger's description of monopoles}\label{Zwanzigersummarysect}
It is well known that the interaction between a pair of electrically and magnetically charged particles leads to a quantized amount of angular momentum to be stored in the electromagnetic field \cite{Dirac:1931kp}.
In \cite{Zwanziger:1972sx} Zwanziger calculated the relativistic form of this angular momentum. For electric and magnetic charges moving along linear trajectories $x^{\mu}(\tau)=\frac{p^{\mu}}{m}\tau+\O(\ln\tau)$, one finds that in addition to the usual spin and orbital angular momentum there is a contribution to the electromagnetic field's angular momentum tensor of the form
\begin{align}
\lim\limits_{t\rightarrow \pm \infty}M^{\mu\nu}(t)&=\sum_{i>j}\eta_i\,\mu_{ij}\frac{\e^{\mu\nu}_{\,\,\,\,\,\,\kappa\lambda}p_i^{\kappa}p_j^{\lambda}}{[(p_i\.p_j)^2-m_i^2m_j^2]^{1/2}}\label{angularmomentumtensor}\\
\mu_{ij}&=\frac{e_ig_j-g_ie_j}{4\pi}
\end{align}
where the pairing $i,j$ is only for pairs that are both in or both out states, $\eta_i=+/-$ is used for in/out states respectively, and $e_i,g_j$ refer to the electric and magnetic charges of the particles. At intermediate stages of the calculation leading to (\ref{angularmomentumtensor}), factors of $t$ canceled amongst one another leading to the time independent result. One can then trivially take the late time limit and observe that the asymptotic angular momentum does not coincide with the free theory's angular momentum. This suggests that in the quantum theory the asymptotic states do not transform as free particle representations. To better understand the relativistic form (\ref{angularmomentumtensor}), consider the case of two particles in a frame where one of them is at rest. In this frame there is no boost charge $M^{0i}=0$, but there is an angular momentum $\vec{J}=-\eta_i\mu_{ij}\hat{n}_j$ ($J_i=\frac{1}{2}\e_{ijk}M^{jk}$, and $\hat{n}_j=\vec{p}_j/E_j$). Similar to our case of the boost charge in Section \ref{sectboostcharge} we see a Lorentz charge associated to the direction of motion of a charged particle. Again this hints at why the celestial sphere representations may be an appropriate basis for asymptotically interacting representations: a $\ket{z;\triangle,\sigma}$ state is an eigenstate of rotations about it's stereographic direction $z$, with eigenvalue proportional to it's spin $\sigma$.\par
We now outline how Zwanziger extended Wigner's representations of the Poincar\'e group to include asymptotic electric-magnetic interactions. First observe that for any two momenta $(p_1^{\mu},p_2^{\mu})$, which are not collinear, one can always Lorentz transform to the center-of-momentum (COM) frame where the momenta point along the $z$ axis
\begin{gather}
\tilde{k}_1=(\tilde{E}_1,0,0,\tilde{p}_c),\quad \tilde{k}_2=(\tilde{E}_2,0,0,-\tilde{p}_c)\label{refstate}\\
\tilde{E}_{1,2}=\sqrt{m_{1,2}^2+p_c^2},\quad \tilde{p}_c=\sqrt{\frac{(p_1\.p_2)^2-m_1^2m_2^2}{s}}.
\end{gather}
In this frame the only non-zero component of the classical angular momentum (\ref{angularmomentumtensor}) is
\begin{equation}
(M^{xy}_{1,2})_{\text{C.O.M.}}=\eta_1 \mu_{12}.\label{rotationcharge}
\end{equation}
 By analogy to Wigner's construction of representations of the Poincar\'e group, one takes (\ref{refstate}) to be a reference state from which all other two particle states are defined. Let $L(p_1,p_2)$ denote a Lorentz transformation which simultaneously carries $\tilde{k}_1^{\mu}$ to $p_1^{\mu}$ and $\tilde{k}_2^{\mu}$ to $p_2^{\mu}$. We then define two particle states at arbitrary momenta by a standard boost $L(p_1,p_2)$ applied to the reference state
\begin{equation}
\ket{p_1,p_2;q_{12}}\defined U[L(p_1,p_2)]\ket{\tilde{k}_1,\tilde{k}_2;q_{12}},\label{defoftwoparticle}
\end{equation}
where $q_{12}$ denotes any internal quantum numbers of the pairwise state, the nature of which will be determined shortly. The definition (\ref{defoftwoparticle}) fixes the definition of the phase and quantum numbers of a pairwise state at a given pair of momenta, thus allowing one to compare two particle states with the same momenta but different quantum numbers. One can now compute the effect of a Lorentz transformation. For simplicity we will first consider the case of scalar particles. The modification required to account for the spin and helicity of particles is simple and will be explained thereafter. To determine the effect of a Lorentz transformation, one employs the standard set of manipulations used in the method of induced representations \cite{Wigner,Weinberg:1995mt}:
\begin{align*}
U[\Lambda]&\ket{p_1,p_2;q_{12}}\\
&\,\,\defined U[\Lambda]U[L(p_1,p_2)]\ket{\tilde{k}_1,\tilde{k}_2;q_{12}}\label{def}\numberthis\\
&\,\,=U[L(\Lambda p_1,\Lambda p_2)]\bigg(U[L^{-1}(\Lambda p_1,\Lambda p_2)]U[\Lambda]U[L(p_1,p_2)]\bigg)\ket{\tilde{k}_1,\tilde{k}_2;q_{12}}\numberthis\label{identity}\\
&\,\,=U[L(\Lambda p_1,\Lambda p_2)]e^{-iq_{12}\theta_{12}}\ket{\tilde{k}_1,\tilde{k}_2;q_{12}}\numberthis\label{commutestep}\\
&\,\,=e^{-iq_{12}\theta_{12}}\ket{\Lambda p_1,\Lambda p_2;q_{12}}.\numberthis\label{finalLorentz}
\end{align*}
Line (\ref{def}) follows from the definition of the pairwise state (\ref{defoftwoparticle}).
In going to line (\ref{identity}) one inserts the identity operator in the form 
\begin{equation}
\I=U[L(\Lambda p_1,\Lambda p_2)]U[L^{-1}(\Lambda p_1,\Lambda p_2)].
\end{equation} 
One then recognizes that the term in brackets in (\ref{identity}) must be an element of the little groups of both $\tilde{k}_1$ and $\tilde{k}_2$. To see this, note that the term in big brackets carries the reference momenta along the path
\begin{equation}
(\tilde{k}_1,\tilde{k}_2)\,\,\overset{L(p_1,p_2)\,}{\longrightarrow}\,\, (p_1,p_2)\,\,\overset{\Lambda}{\longrightarrow}\,\, (\Lambda p_1,\Lambda p_2)\,\,\overset{L^{-1}(\Lambda p_1,\Lambda p_2)\,\,\,}{\longrightarrow}\,\,(\tilde{k}_1,\tilde{k}_2).
\end{equation}
The intersection of the little groups of $\tilde{k}_1$ and $\tilde{k}_2$ is the group of rotations about the $z$-axis. This group is abelian and consequently it's unitary irreducible representations are represented by a pure phase. We denote this phase by $e^{-iq_{12}\,\,\theta_{12}}$, where we have omitted the Wigner phase's dependence on the momenta and Lorentz transformation $\theta_{12}(\Lambda; p_1,p_2;\tilde{k}_1,\tilde{k}_2)$ for brevity. Comparing to the classical angular momentum stored in the electromagnetic field in the COM frame (\ref{rotationcharge}), it is natural to assign the quantum number $q_{12}$ the value
\begin{equation}
q_{12}=\frac{e_1g_2-e_2g_1}{4\pi}\defined\mu_{12}.\label{electricmagneticcharge}
\end{equation}
Altogether we arrive at the pairwise transformation law
\begin{equation}
U[\Lambda]\ket{p_1,p_2}=e^{-i\mu_{12}\theta_{12}}\ket{\Lambda p_1,\Lambda p_2}\label{transform1}.
\end{equation}
The method of induced representations guarantees that (\ref{transform1}) does indeed form a representation of the Poincar\'e group, and furthermore it is manifestly unitary. 
The topology of the $U(1)$ little group restricts the charge $\mu_{12}$ to be quantized, which is the familiar Dirac-Zwanziger result that the product of the electric and magnetic charge must take integer values.  To include the spin and helicity of the particles one acts with the usual $D_{\sigma\sigma'}[W(\Lambda,p)]$ Wigner matrices
\begin{equation}
U[\Lambda]\ket{p_1,\sigma_1;p_2,\sigma_2}=e^{-i\mu_{12}\theta_{12}}D_{\sigma_1\sigma_1'}D_{\sigma_2\sigma_2'}\ket{\Lambda p_1,\sigma_1';\Lambda p_2,\sigma_2'},
\end{equation} 
with the justification that this again forms a representation, and furthermore there is no reason to suspect that asymptotic electric-magnetic interactions spoil the helicity and spin quantum numbers. To generalize to the situation of more than two particles, we note that the classical modification to the angular momentum tensor (\ref{angularmomentumtensor}) is a sum of terms that involves two particles at a time. This suggests that the multi-particle transformation law involves a Zwanziger little group phase for each electric-magnetic pair of particles. Let $\ket{\set{p,\sigma;e,g}}\defined\prod_{\otimes_i}\ket{p_i,\sigma_i;e_i,g_i}$ denote the in state. Then the multi-particle transformation law is
\begin{equation}
U[\Lambda]\ket{\set{p,\sigma;e,g}}=\bigg(\prod_{i<j}e^{-i\mu_{ij}\theta_{ij}}\bigg)\bigg(\prod_i D_{\sigma_i\sigma_i'}\bigg)\ket{\set{\Lambda p,\sigma;e,g}}.\label{multiparticle}
\end{equation}
The only difference between (\ref{multiparticle}) and the usual transformation law is the first term in brackets, which takes into account the late time dynamics of the electric-magnetic charge pairs.  To complete the discussion of representations of the Poincar\'e group we must decide how the translation generators act on the pairwise state. In \cite{Zwanziger:1972sx}, Zwanziger calculates the classical energy-momentum tensor of the electromagnetic field for electric-magnetic charges moving along linear trajectories and found that their contribution vanishes at asymptotic times.  This motivates us to declare that the translation generators $\hat{P}^{\mu}$ act in the usual manner
\begin{equation}
\hat{P}^{\mu}\ket{\set{p,\sigma;e,g}}=\bigg(\sum_i p_i^{\mu}\bigg)\ket{\set{p,\sigma;e,g}}\label{momentum}
\end{equation} 
 with the further justification that (\ref{multiparticle}) and (\ref{momentum}) consistently combine to form a unitary representation of the Poincar\'e group.  \par
Let us now emphasize a conceptual point which was first developed in \cite{Csaki:2020inw,Csaki:2020uun}.  For simplicity let us return to the proposed transformation law for a two particle state
\begin{equation}
U[\Lambda]\ket{p_i,\sigma_i,p_j,\sigma_j;\mu_{ij}}=e^{-i\mu_{ij}\theta_{ij}}D_{\sigma_i\sigma_i'}D_{\sigma_j\sigma_j'}\ket{\Lambda p_i,\sigma_i',\Lambda p_j,\sigma_j';\mu_{ij}}.\label{simpletwo}
\end{equation}
We can view this representation in two different ways; either as a generalization of the tensor product between two representations, or as the ordinary tensor product between \textit{three} representations. We find the latter approach to be the more beneficial for our purposes. That is, we consider (\ref{simpletwo}) to be the result of a Lorentz transformation acting on the triple tensor product state
\begin{equation}
\ket{p_i,\sigma_i,p_j,\sigma_j;\mu_{ij}}=\ket{p_i,\sigma_i}\otimes\ket{p_j,\sigma_j}\otimes \ket{\set{p_i,p_j;\mu_{ij}}}.\label{tripletensor}
\end{equation}
The braces notation used for the last state is used to distinguish this state from an ordinary tensor product of two states, and we refer to this state as the pairwise state. In order to reproduce the transformation laws (\ref{momentum}) and (\ref{simpletwo})  we declare the Poincar\'e transformation properties of the pairwise state to be
\begin{align}
\hat{P}^{\mu}\ket{\set{p_i,p_j;\mu_{ij}}}&=0\label{innerstatenull}\\
U[\Lambda]\ket{\set{p_i,p_j;\mu_{ij}}}&=e^{-i\mu_{ij}\theta_{ij}}\ket{\set{\Lambda p_i,\Lambda p_j;\mu_{ij}}},
\end{align}
and then the other two states in (\ref{tripletensor}) can assume their free theory transformation properties
\begin{align}
\hat{P}^{\mu}\ket{p,\sigma}&=p^{\mu}\ket{p,\sigma}\\
U[\Lambda]\ket{p,\sigma}&=D_{\sigma\sigma'}\ket{\Lambda p,\sigma'}.
\end{align}
We note that this is not the only consistent manner in which one can distribute the Poincar\'e transformation properties amongst the three states, although we find our choice to be the most natural one in light of Zwanziger's calculation that there is no momentum stored in the electromagnetic field at asymptotic times. We refer the interested reader to \cite{Csaki:2020uun} for further discussion on this point. \par
If we adopt the viewpoint that the transformation law (\ref{simpletwo}) arises from acting on a triple tensor product of states, then we see that the state (\ref{simpletwo}) is not a new representation of the Poincar\'e group, but instead just a specific combination of known representations. The virtue then of Zwanziger's prescription is that it singles out a particular choice for the third representation in (\ref{tripletensor}). A further advantage of Zwanziger's prescription is that it puts the electric and magnetic charges in an exponent. The amplitudes consistent with the modified transformation law will then also share this feature.  This then indicates the non-perturbative, all orders in the coupling, nature of Zwanziger's prescription. We therefore observe that the traditional perturbation series, where one works order by order in the coupling, obscures the underlying covariance of the asymptotic states.  
\subsection{Representation theory of celestial amplitudes}\label{sectsummaryofcelestial}
In the foundational work \cite{Strominger:2013lka}, and then further established in \cite{A1Strominger:2013jfa,A2He:2014laa,A3Kapec:2014opa,A4He:2014cra}, it was demonstrated that Weinberg's soft theorems \cite{Weinberg:1965nx,Weinberg:1995mt} are the Ward identities associated to the asymptotic symmetries of gauge theories and gravity. Furthermore, it was found therein that these Ward identities find their most natural expression in a form that resembles a $2$D conformal field theory. This has lead to an expectation that the asymptotic symmetries of gauge theory and gravity may be better elucidated not in the usual four-momentum basis but instead in the so-called celestial sphere basis \cite{c2Lam:2017ofc,c3Schreiber:2017jsr,c4Stieberger:2018edy,c5Stieberger:2018onx,c6Fan:2019emx,c7Pate:2019mfs,c8Puhm:2019zbl,c9Nandan:2019jas,c10Adamo:2019ipt,c11Guevara:2019ypd,c12Pate:2019lpp,c13Law:2019glh,c14Donnay:2020guq,c15Albayrak:2020saa,c16Casali:2020vuy,c165Casali:2020uvr,c18Kapec:2014opa,c19He:2014cra,c20Strominger:2014pwa,c21He:2015zea,c22Pasterski:2015zua,c23Muck:2020wtx,c24Narayanan:2020amh,c25Banerjee:2020zlg,c26Gonzalez:2020tpi,c27Cachazo:2014dia,c30Choi:2017ylo,c31Donnay:2018neh,c32StromLec}, which more closely exhibits a $2$D CFT structure. In this section we summarize the representation theory of the celestial sphere basis. We will restrict our discussion to massless representations, as the massive case is beyond the scope of this paper. We refer the interested reader to \cite{Pasterski:2016qvg,Pasterski:2017kqt,Banerjee:2018gce,Gelfand} for a more complete discussion on the representation theory of celestial states.  \par
We present two routes leading to the celestial sphere representation of amplitudes. The first is by a Mellin transformation on the energy variable of an ordinary Wigner state \cite{Pasterski:2016qvg,Pasterski:2017kqt}, and the second is by inducing from the so-called lower triangular subgroup of the Lorentz group \cite{Banerjee:2018gce}. The Mellin transform approach emphasizes the physical relevance of celestial amplitudes---they are just a change of basis from ordinary amplitudes. The method of inducing from the lower triangular group is less direct, however it is the approach which most readily allows us to generalize Zwanziger's pairwise little group to include boosts and late time electric-electric interactions. In Section \ref{mellinsect} we briefly review the Mellin transform approach. In Section \ref{lowertriangularsect} we summarize the method of inducing from the lower triangular subgroup. 
\subsubsection{A Mellin transform of ordinary amplitudes}\label{mellinsect}
Let $\ket{p^{\mu},\sigma}$ denotes a massless state with helicity $\sigma$. We will parameterize the massless momenta using stereographic co-ordinates
\begin{gather}
p^{\mu}(E,z)=\frac{E}{1+|z|^2}\bigg(1+|z|^2,z+\bar{z},-i(z-\bar{z}),1-|z|^2\bigg)\\
z=e^{i\phi}\tan\frac{\theta}{2}.\label{stereoz}
\end{gather}
We then define the celestial state $\ket{z;\triangle,\sigma}$ as the Mellin transform with respect to energy variable of the Wigner state
\begin{equation}
\ket{z;\triangle,\sigma}\defined\bigg(\frac{1}{1+|z|^2}\bigg)^{i\eta\lambda+1}\int_{0}^{\infty}\DD E\,\,\, E^{\,i\eta\lambda}\ket{p^{\mu},\sigma}.\label{ordinarymellin}
\end{equation}
\footnote{In some works, for example \cite{Pasterski:2016qvg}, the $(1+|z|^2)^{i\eta \lambda+1}$ prefactor is not explicitly indicated. In those works the energy variable with respect to which one is taking the Mellin transform is understood to be $\omega= \frac{E}{1+|z|^2}$.}where $\eta=+/-$ for in/out states respectively. The parameter $\triangle=1+i\lambda$ is referred to as the conformal dimension of the state, whose namesake will become apparent shortly. One can then make use of the known transformation property of the $\ket{p^{\mu},\sigma}$ states, $U[\Lambda]\ket{p,\sigma}=e^{-i\sigma \theta}\ket{\Lambda p,\sigma}$, to determine the transformation property of the celestial states. What one finds is that\footnote{Throughout this paper we use the Mobius representation of $SL(2,\mathbb{C})$ as co-ordinates for the Lorentz group manifold:
\begin{equation}
\Lambda\in \begin{pmatrix}
a & b\\
c & d
\end{pmatrix},\quad ad-bc=1,\quad a,b,c,d\in\mathbb{C}.
\end{equation}}
\begin{align}
U[\Lambda]\ket{z;\triangle,\sigma}=|cz+d|^{-2i\eta\lambda-2}\bigg(\frac{cz+d}{\bar{c}\bar{z}+\bar{d}}\bigg)^{\eta\sigma}\bigg\vert \frac{az+b}{cz+d};\triangle,\sigma\bigg\rangle.\label{cprimary}
\end{align}
So we see that every Lorentz transformation $\Lambda$ induces a Mobius transformation on the stereographic direction $z=e^{i\phi}\tan\frac{\theta}{2}$ of a massless particle.  Another interesting feature of the celestial states is that they are eigenstates of boosts in their direction of motion; a boost in the $z=e^{i\phi}\tan\frac{\theta}{2}$ direction will leave $z$ invariant, hence (\ref{cprimary}) indicates that $\ket{z;\triangle,\sigma}$ is an eigenstate of such boosts with an eigenvalue proportional to it's conformal dimension $\triangle$.  Structurally (\ref{cprimary}) is also the transformation law for a quasi-primary state in a two dimensional Euclidean theory with global conformal invariance, with  holomorphic/anti-holomorphic conformal dimension $(h,\bar{h})=(\frac{\triangle-\sigma}{2},\frac{\triangle+\sigma}{2})$. The resemblance to a $2$D CFT is a consequence of the isomorphism between the Lorentz group and the global conformal group of two dimensional Euclidean space.  This provides another motivation for studying this basis; the structural similarity to a $2$D CFT may be the first indications of a possible holographic description of flat space amplitudes \cite{deBoer:2003vf}, where all Minkowski space amplitudes are described by an as of yet unknown $2$D CFT.\par
 The reader may also recognize (\ref{cprimary}) to be the principal series representation of the Lorentz group \cite{Gelfand,Naimark}.  Therein it was established that such representations (\ref{cprimary}) are unitary with respect to the inner product
 \begin{equation}
 \braket{z_2\triangle_2,\sigma_2|z_1,\triangle_1,\sigma_1}=\delta^{2}(z_1-z_2)\delta(\lambda_1-\lambda_2)\delta_{\sigma_2\sigma_1},
 \end{equation} with the restriction that the conformal dimension and helicity must assume either the form $\lambda\in \mathbb{R}\,,\,\sigma\in \frac{\mathbb{Z}}{2}$ (principal series) or $\lambda= ic\, ,\,\sigma=0$ with $0\leq|c|\leq1,\,\, c\in \mathbb{R}$ (supplementary series) in order for the representation to be unitary.  See \cite{Pasterski:2017kqt,Gelfand} for a more complete discussion of how unitarity restricts the values of the conformal dimension and helicity parameters.\par
 To convert a Minkowski amplitude $\A$ to a celestial amplitude $\A^{\text{C.S.}}$ we Mellin transform in the same manner as in (\ref{ordinarymellin}) 
\begin{equation}
\A^{\text{C.S.}}(\set{z;\triangle,\sigma})=\bigg(\prod_j \bigg(\frac{1}{1+|z_j|^2}\bigg)^{i\eta_j\lambda_j+1}\int_0^{\infty}\DD E_j\,\, E_j^{\,i\eta_j\lambda_j}\bigg)\A(\set{p^{\mu},\sigma}),\label{convertMinktoCelest}
\end{equation} 
where $\set{z;\triangle,\sigma}$ and $\set{p^{\mu},\sigma}$ are shorthands for the set of all particle labels, and $\eta_j=+/-$ is used for in/out states respectively. The $SL(2,\mathbb{C})$ covariance of the celestial states (\ref{cprimary}) then translates into the $SL(2,\mathbb{C})$ covariance of the celestial amplitudes as
\begin{equation}
\A^{\text{C.S.}}\big(\set{z;\triangle,\sigma}\big)=\bigg(\prod_j|cz_j+d|^{-2i\eta_j\lambda_j-2}\bigg(\frac{cz_j+d}{\bar{c}\bar{z}_j+\bar{d}}\bigg)^{\eta_j\sigma_j}\bigg)\A^{\text{C.S.}}\bigg(\bigg\lbrace \frac{az+b}{cz+d};\triangle,\sigma\bigg\rbrace\bigg).\label{amplitudecovariance}
\end{equation} 
 We will make use of (\ref{amplitudecovariance}) in Section \ref{sectamplitudes} when constructing $SL(2,\mathbb{C})$ covariant celestial amplitudes. 
\subsubsection{Inducing from the Lower Triangular Subgroup of the Lorentz Group}\label{lowertriangularsect}
In \cite{Banerjee:2018gce}, the celestial sphere representation (\ref{cprimary}) was derived using the method of induced representations.  In this approach, the Lorentz subgroup which one induces from is $\set{J_3,K_3, J_2-K_1,J_1+K_2}$ which is the usual little group for a massless particle traveling along the third direction, but with the addition of boosts in the third direction $K_3$. It is this enhancement of the little group that occurs for celestial states that will allow us to also enhance Zwanziger's pairwise little group in Section \ref{sect-asym,int,cels,rep}. In this section we summarize the construction in \cite{Banerjee:2018gce}.\par
 We begin with the following representation of the Lorentz group
\begin{equation}
U[\Lambda]\ket{z}=\bigg\vert\frac{az+b}{cz+d}\bigg\rangle.\label{startrep}
\end{equation}
The reader can readily verify that this is indeed a representation. Prior to declaring an inner product, one cannot yet make any claim as to whether the representation is unitary.  It will be expedient to introduce the shorthand notation
\begin{equation}
\frac{az+b}{cz+d}\defined \Lambda z.
\end{equation} We will now enhance the representation (\ref{startrep}) by making use of the method of induced representations. We therefore add an additional set of quantum numbers $q$ to our state, the nature of which will be determined shortly. We will take the $z=0$ state as our reference state, from which all other states are defined via a standard boost $L(z)$
\begin{equation}
\ket{z,q}=N(z)U[L(z)]\ket{0;q},\label{defofonecelest}
\end{equation}
 \footnote{According to (\ref{startrep}), the $z=0$ direction gets mapped to $z=\frac{b}{d}$, so the standard boost is any $L(z)={\tiny\begin{pmatrix}
 a & b\\
 c& d
 \end{pmatrix}}\in SL(2,\mathbb{C})$ with the restriction that $\frac{b}{d}=z$. }where $N(z)$ is a normalization factor that will be chosen at the end of the calculation to simplify the result. To determine the effect of a Lorentz transformation we employ the standard set of manipulations used in the method of induced representations
\begin{align}
U[\Lambda]\ket{z,q}&=N(z) U[\Lambda]U[L(z)]\ket{0;q}\\
&=N(z)U[L(\Lambda z)]\bigg(U[L^{-1}(\Lambda z)]U[\Lambda]U[L(z)]\bigg)\ket{0;q}\label{resolveiden}\\
&=N(z)U[L(\Lambda z)]W_{q q'}(\alpha,\beta)\ket{0;q'}\label{interm}\\
&=\frac{N(z)}{N(\Lambda z)}W_{qq'}(\alpha,\beta)\ket{\Lambda z;q'}.\label{transformofcelestial}
\end{align}
The justification for these manipulations is the exact same as that given below (\ref{finalLorentz}), so we will only discuss the nature of the little group, whose representation we have denoted as $W_{qq'}(\alpha,\beta)$ in (\ref{interm}). Recognize that the term in brackets in (\ref{resolveiden}) carries $z=0$ along the following path
\begin{equation}
0\,\,\overset{L(z)}{\longrightarrow}\,\,z\,\,\overset{\Lambda}{\longrightarrow }\,\,\Lambda z\,\,\overset{L^{-1}(\Lambda z)}{\longrightarrow }0.
\end{equation} 
We therefore see that the term in brackets in (\ref{resolveiden}) must be an element of the little group of the $z=0$ direction. We can readily determine the nature of this little group by observing that a Lorentz transformation acts on the $z=0$ direction as
\begin{equation}
\Lambda \,\,\colon \,\,0\mapsto \frac{az+b}{cz+d}\at{z=0}=\frac{b}{d}.
\end{equation} 
So we see that the little group of $z=0$ is the set of Lorentz transformations for which $b=0$ and $d$ is finite. Accounting for the $ad-bc=1$ condition we can parameterize the little group $\text{LG}_{z=0}$ as the following set of $2\times 2$ matrices
\begin{equation}
\text{LG}_{z=0}=\bigg\lbrace \begin{pmatrix}
a & 0\\
c & a^{-1}
\end{pmatrix},\,\,\, a,c\in\mathbb{C}\label{paralittle}\bigg\rbrace.
\end{equation} 
Hence we are inducing from the so-called lower triangular subgroup of the Lorentz group. This subgroup is generated by the Lie algebra elements 
\begin{equation}
\set{J_3,K_3,J_2-K_1,J_1+K_2},\label{setoflittlegroupgenerators}
\end{equation}
which is the ordinary little group of a massless particle appended by $K_3$. As emphasized in \cite{Banerjee:2018gce}, this the group that leaves the $z=0$ stereographic \textit{direction} invariant, as opposed to the $p^{\mu}=(E,0,0,E)$ momentum invariant. We will assume that our representation does not contain any continuous spin particles, that is we will declare that our reference state is annihilated by the following two little group generators
\begin{equation}
(J_2-K_1)\ket{0;q}=(J_1+K_2)\ket{0;q}=0.\label{annihilateCSR}
\end{equation} 
As the remaining elements of the little group commute $[K_3,J_3]=0$, we can simultaneously diagonalize these operators and label our states by two quantum numbers $q=\triangle,\sigma$ where
\begin{equation}
J_3\ket{0;\triangle,\sigma}=\sigma\ket{0;\triangle,\sigma},\qquad K_3\ket{0;\triangle,\sigma}=-i\triangle\ket{0;\triangle,\sigma},\label{eigenref}
\end{equation}
where we chose the boost eigenvalue to take the form $-i\triangle$ so that our final result will match our use of $\triangle$ in Section \ref{mellinsect}\footnote{Note that we will be setting $\triangle=1+i\lambda,\lambda\in\mathbb{R}$, in which case (\ref{eigenref}) indicates that the $K_3$ operator has complex eigenvalues. This does not contradict the fact that $K_3$ is Hermitian, because it is not self-adjoint \cite{Gieres:1999zv}. The requirement that the representation be unitary only requires that $K_3$ be Hermitian, and we do not need the stronger condition that the operator be self-adjoint.}. In this basis the action of the little group is one dimensional, that is the quantum numbers $\triangle,\sigma$ do not change. Returning to our transformation law (\ref{transformofcelestial}), we replace $q\rightarrow \triangle,\sigma$ and insert our representation of the little group acting on the reference state $W(\alpha,\beta)\ket{0;\triangle,\sigma}=e^{-i\alpha J_3}e^{-i\beta K_3}\ket{0;\triangle,\sigma}=e^{-i\sigma \alpha }e^{-\triangle \beta}\ket{0;\triangle,\sigma}$, which altogether gives
\begin{align}
U[\Lambda]\ket{z;\triangle,\sigma}=\frac{N(z)}{N(\Lambda z)}e^{-i \sigma \alpha}e^{- \triangle\beta}\ket{\Lambda z;\triangle,\sigma},\label{intermedietestepforinducingfromlowertriangular}
\end{align}
where we have omitted the little group parameters dependence $\alpha(z,\Lambda),\beta(z,\Lambda)$ for brevity. One can solve for these little group parameters explicitly. As we perform a very similar calculation in Section \ref{sect-asym,int,cels,rep} we will only quote the result\footnote{These parameters depend on one's choice of standard boost. The results quoted at (\ref{littlegroupparametersexplictly1}, \ref{littlegroupparametersexplictly2}) result from the choice $L(z)={\tiny\begin{pmatrix}
1 & \frac{z}{1-z}\\
1 &\frac{1}{1-z}
\end{pmatrix}}$.}
\begin{gather}
e^{-i\alpha(z,\Lambda)}=\frac{c z+d}{\bar{c}\bar{z}+\bar{d}}\,\bigg(\frac{1-\bar{z}}{1-z}\bigg)\bigg(\frac{1-\Lambda z	}{1-\Lambda \bar{z}}\bigg)\label{littlegroupparametersexplictly1}\\
 e^{-i\beta(z,\Lambda)}=|cz+d|^{-2i}\,\bigg|\frac{1-z}{1-\Lambda z}\bigg|^{2i}.\label{littlegroupparametersexplictly2}
\end{gather}
In order to simplify our final result we make the following choice for the normalization
\begin{equation}
N(z)=\bigg(\frac{1-z}{1-\bar{z}}\bigg)^{\sigma}|1-z|^{-2\triangle}\label{norm}.
\end{equation}
Then combining (\ref{intermedietestepforinducingfromlowertriangular}, \ref{littlegroupparametersexplictly1}, \ref{littlegroupparametersexplictly2}, \ref{norm}) we arrive at our final form for the transformation law of the celestial state
\begin{equation}
U[\Lambda]\ket{z;\triangle,\sigma}=|cz+d|^{-2\triangle}\bigg(\frac{c z+d}{\bar{c}\bar{z}+\bar{d}}\bigg)^{\sigma}\ket{\Lambda z;\triangle,\sigma}\label{celesttransform}.
\end{equation}
Then if we identify $\triangle=1+i\lambda$, and add the appropriate $\eta=+/-$ factors to distinguish in/out states, we see that (\ref{celesttransform}) is exactly the same transformation law as in (\ref{cprimary}). In summary, one can construct the principal series representations (\ref{cprimary}, \ref{celesttransform}) by inducing from the (\ref{paralittle}) subgroup of the Lorentz group which leaves the $z=0$ direction invariant.
\section{Asymptotically interacting celestial representations}\label{sect-asym,int,cels,rep}
In this section we demonstrate that Zwanziger's pairwise little group is enhanced to also include a boost subgroup when applied to celestial states. In Section \ref{sectcompare} we will demonstrate that this boost subgroup correctly accounts for the asymptotic dynamics in abelian gauge theories containing massless charged particles. The germ of the idea is easy to appreciate; two celestial states, one pointing along the north pole $z=0$, the other along the south pole $z=\infty$, have a pairwise little group generated by the elements $\set{J_3,K_3}$, as both of these generators leave both the north and south pole directions invariant. Zwanziger showed that the $J_3$ pairwise little group accounts for the asymptotic electric-magnetic interactions $e_ig_j$, and we will demonstrate that the $K_3$ pairwise little group accounts for the late time electric-electric $e_ie_j$ interactions\footnote{It is possible that the $K_3$ little group may also correctly account for purely magnetic $g_ig_j$ late time interactions, although we do not explore this possibility in the present paper. }.  In Section \ref{sectcelestialzwanziger} we describe the pairwise transformation law for two celestial states. In Section \ref{sectamplitudes} we construct amplitudes consistent with the modified transformation law. 
\subsection{The pairwise little group of celestial states}\label{sectcelestialzwanziger}
Let us begin with the tensor product of two celestial states
\begin{equation}
\ket{z_1;\triangle_1,\sigma_1}\otimes\ket{z_2;\triangle_2,\sigma_2}.
\end{equation}
Following the discussion around (\ref{tripletensor}) we will now enhance this representation to a triple tensor product
\begin{equation}
\ket{z_1;\triangle_1,\sigma_1}\otimes\ket{z_2;\triangle_2,\sigma_2}\otimes\ket{\set{z_1,z_2;q}},\label{celesttriple}
\end{equation}
where the first two states in (\ref{celesttriple}) transform as ordinary celestial representations (\ref{cprimary}, \ref{celesttransform}).  The modification to the transform law will be entirely contained in the last state $\ket{\set{z_1,z_2;q}}$, which we will refer to as the pairwise celestial state. We use braces notation $\set{}$ for the pairwise celestial state to distinguish it from an ordinary celestial state. The rest of this subsection is dedicated to determining the nature of the quantum numbers $q$ and the transformation properties of the pairwise celestial state. As such we will drop the first two terms in (\ref{celesttriple}), and restore their presence at the end of the calculation. \par
Given two stereographic directions $z_1,z_2$ we can always Lorentz transform ${\tiny\begin{pmatrix}
a & b \\
c & d
\end{pmatrix}}$ to a frame where $z_1$ points along the north pole $z=0$ and $z_2$ points along the south pole $z=\infty$\footnote{Barring the one exception where the particles are collinear $z_1=z_2$.}. The converse of this is that we may therefore express any two particle state as a standard boost $L(z_1,z_2)$ applied to the north-south reference state
\begin{equation}
\ket{\set{z_1,z_2;q}}\defined U[L(z_1,z_2)]\ket{\set{0,\infty;q}}\label{standardboost} .
\end{equation}
Here we could add a possible normalization factor $N(z_1,z_2)$, however by the end of the calculation we would conclude that $N=1$ is the most suitable choice, so we will omit it. By using the same quantum numbers $q$ on either side of (\ref{standardboost}), this equation fixes the phase and internal quantum numbers $q$ at a given pair of stereographic directions, allowing us to compare two pairwise celestial states with the same stereographic directions but different quantum numbers. 
In order to determine the effect of a Lorentz transformation, we employ the standard set of manipulations used in the method of induced representations \cite{Wigner,Weinberg:1995mt}:
\begin{align*}
U[\Lambda]&\ket{\set{z_1,z_2;q}}\\
&=U[\Lambda]U[L(z_1,z_2)]\ket{\set{0,\infty;q}}\label{secondline}\numberthis\\
&=U[L(\Lambda z_1,\Lambda z_2)]\bigg(U[L^{-1}(\Lambda z_1,\Lambda z_2)]U[\Lambda]U[L(z_1,z_2)]\bigg)\ket{\set{0,\infty;q}}\numberthis\label{brackets}\\
&=D_{qq'}(\alpha,\beta)U[L(\Lambda z_1,\Lambda z_2)]\ket{\set{0,\infty;q'}}\label{intermedstepofpairwisecelest}\numberthis\\
&=D_{qq'}(\alpha,\beta)\ket{\set{\Lambda z_1,\Lambda z_2;q'}}\label{finalpairwise}\numberthis .
\end{align*}
The justification for these manipulations are the exact same as those given below (\ref{finalLorentz}), but we will repeat their explanation here for completeness.  Line (\ref{secondline}) follows from the definition (\ref{standardboost}) of the pairwise celestial state. In going to line (\ref{brackets}) we have inserted the identity operator in the form
\begin{equation}
\I=U[L(\Lambda z_1,\Lambda z_2)]U[L^{-1}(\Lambda z_1,\Lambda z_2)].
\end{equation} 
Then we recognize that the term in brackets in (\ref{brackets}) must simultaneously be an element of the little groups of both the north $z=0$ and south $z=\infty$ directions. To see this, notice that the term in brackets carries the reference directions along the following path: 
\begin{equation}
(0,\infty)\,\,\overset{L(z_1,z_2)\,}{\longrightarrow}\,\, (z_1,z_2)\,\,\overset{\Lambda}{\longrightarrow}\,\, (\Lambda z_1,\Lambda z_2)\,\,\overset{L^{-1}(\Lambda z_1,\Lambda z_2)\,\,\,}{\longrightarrow}\,\,(0,\infty).
\end{equation}
 The intersection of the little groups of the north pole and south pole directions is the group of boosts along and rotations about the third axis, as both of these operations leave both the north and south pole directions invariant. This subgroup is two dimensional and hence in (\ref{intermedstepofpairwisecelest}) we denote the representation of the little group acting on the internal quantum numbers $q$, as $D_{qq'}(\alpha,\beta)$, where we omit the Wigner parameters dependence $\alpha(\Lambda;z_1,z_2),\beta(\Lambda;z_1,z_2)$ for brevity. The final line (\ref{finalpairwise}) then follows again from the definition (\ref{standardboost}).\par
   As the elements of the little group's algebra commute $[J_3,K_3]=0$, we may simultaneously diagonalize these operators on our representation space. We therefore take our pairwise reference state to be eigenstates of these operators with eigenvalues $\sigma_{12},\triangle_{12}$
 \begin{align}
 \hat{J}_3\ket{\set{0,\infty;\sigma_{12},\triangle_{12}}}&=\sigma_{12}\ket{\set{0,\infty;\sigma_{12},\triangle_{12}}}\label{roteign}\\
 \hat{K}_3\ket{\set{0,\infty;\sigma_{12},\triangle_{12}}}&=\triangle_{12}\ket{\set{0,\infty;\sigma_{12},\triangle_{12}}}.\label{boosteign}
 \end{align} 
 We refer to these quantum numbers as the \textit{pairwise} helicity $\sigma_{12}$ and \textit{pairwise} conformal dimension $\triangle_{12}$. We expect that for the case of QED the $\sigma_{12}$ charge accounts for electric-magnetic interactions $e_1g_2-e_2g_1$ and that $\triangle_{12}$ charge accounts for electric-electric and magnetic-magnetic interactions $e_1e_2+g_1g_2$\footnote{We have not verified this last claim that $\triangle_{12}$ may also account for magnetic-magnetic interactions, however the formalism here strongly suggests this to be the case.}. Again, Zwanziger's observation \cite{Zwanziger:1972sx} applies, that the topology of the $J_3$ little group restricts the $\sigma_{12}$ quantum number to take integer multiple values of $\frac{1}{4\pi}$, which is the familiar Dirac-Zwanziger quantization condition of electric-magnetic charge. There is no such quantization condition for $\triangle_{12}$ as the $K_3$ little subgroup is topologically $\mathbb{R}^1$.  Using (\ref{roteign}, \ref{boosteign}) the action of the little group on the north-south reference state is $1$-dimensional 
\begin{align}
\hat{D}(\alpha,\beta)\ket{\set{0,\infty;\sigma_{12},\triangle_{12}}}&=e^{-i\alpha \hat{J}_3}e^{-i\beta\hat{K}_3}\ket{\set{0,\infty;\sigma_{12},\triangle_{12}}}\\
&=e^{-i\sigma_{12}\alpha }e^{-i\triangle_{12}\beta}\ket{\set{0,\infty;\sigma_{12},\triangle_{12}}}\label{littleonnorthsouthstate}. 
\end{align}
 Returning to our pairwise transformation law (\ref{finalpairwise}), and plugging (\ref{littleonnorthsouthstate}) into that equation, we find that our pairwise state transforms as
\begin{equation}
U[\Lambda]\,\ket{\set{z_1,z_2;\sigma_{12},\triangle_{12}}}=e^{-i\sigma_{12}\alpha}e^{-i\triangle_{12}\beta}\ket{\set{\Lambda z_1,\Lambda z_2;\sigma_{12},\triangle_{12}}}.\label{pairwisecelestwithoutexplicitparameters}
\end{equation} 
We now solve for the explicit dependence of the Wigner parameters $\alpha(\Lambda;z_1,z_2),\beta(\Lambda;z_1,z_2)$. First we need to decide on a standard boost $L(z_1,z_2)$. The conditions that the standard boost carries $z=0,z=\infty$ to $z_1,z_2$ respectively are:
\begin{alignat}{2}
z_1&=\frac{a z +b}{c z+d}\at{z=0}&&=\frac{b}{d}\\
z_2&=\frac{az+b}{cz+d}\at{z=\infty}&&=\frac{a}{c}\\
ad-bc&=1\implies z_2-z_1&&=\frac{1}{cd}.
\end{alignat}
So we see that we can choose our standard boost to be:
\begin{equation}
L(z_1,z_2)=\begin{pmatrix}
g z_2 & \frac{z_1}{g(z_2-z_1)}\\
g & \frac{1}{g(z_2-z_1)}
\end{pmatrix}.\label{standardboostmatrix}
\end{equation}
We are free to choose $g$ to be any complex number, as this amounts to a choice in an initial little group transformation prior to changing the stereographic directions. We will keep the choice of $g$ arbitrary in order to emphasize that it will have no effect on the representation\footnote{At this point we may consider whether one can use the fact that there is always a Lorentz transformation which maps three specified directions $(z_1,z_2,z_3)$ to the three points $(0,1,\infty)$ to construct a ``threewise" representation? In this case the intersection of the little groups is the identity element and hence no non-trivial representations will arise.}. We solve for the Wigner parameters by solving the equation
\begin{equation}
e^{-i\alpha \hat{J}_3}e^{-i\beta \hat{K}_3}=[L(\Lambda z_1,\Lambda z_2)]^{-1}\Lambda L(z_1,z_2)\label{wignereq}
\end{equation}
where all matrices here are understood to be in the Mobius representation of the Lorentz group (one can use any non-trivial representation of the Lorentz group to extract the Wigner parameters). On the LHS of (\ref{wignereq}) we have\footnote{To get all of the signs in the exponents correct note the following. An active rotation about the third axis in the right handed sense should increase the $\phi$ parameter of $z=e^{i\phi}\tan\frac{\theta}{2}$, whereas an active boost in the direction of the third axis should bring $z$ closer to the north pole $z=0$ and hence should diminish the value of $z$. Interpreting (\ref{LHSS}) as a Mobius transformation acting on $z$, we see that the signs in the exponents correctly implements this. }
\begin{align}
e^{-i\alpha \hat{J}_3}e^{-i\beta \hat{K}_3}&=\begin{pmatrix}
e^{\frac{(-\beta+i\alpha)}{2}} & 0\\
0 & e^{\frac{\beta-i\alpha}{2}}
\end{pmatrix}.\label{LHSS}
\end{align}
To compute the RHS of (\ref{wignereq}) we will need to invert $L(\Lambda z_1,\Lambda z_2)$. This is especially simple in the Mobius representation where
\begin{equation}
\begin{pmatrix}
a & b \\
 c& d
\end{pmatrix}^{-1}=\begin{pmatrix}
d & -b\\
-c & a
\end{pmatrix}.
\end{equation}
Using our expression (\ref{standardboostmatrix}) for the standard boost matrix, the RHS of (\ref{wignereq}) then reads:
\begin{align*}
&[L(\Lambda z_1,\Lambda z_2)]^{-1}\Lambda L(z_1,z_2)\\
&=\begin{pmatrix}
\frac{1}{g(\Lambda z_2-\Lambda z_1)} & -\frac{\Lambda z_1}{g(\Lambda z_2-\Lambda z_1)}\\
-g & g \,\,\Lambda z_2
\end{pmatrix}\begin{pmatrix}
a & b \\
c & d
\end{pmatrix}\begin{pmatrix}
g z_2 & \frac{z_1}{g(z_2-z_1)}\\
g & \frac{1}{g(z_2-z_1)}
\end{pmatrix}\numberthis\\
&=\begin{pmatrix}
cz_2+d & 0\\
0 & \frac{1}{c z_2+d}
\end{pmatrix}\numberthis\label{RHSS}
\end{align*}
where we note that the $g$ dependence dropped out upon simplifying. Equating (\ref{LHSS}) to (\ref{RHSS}) we find the explicit form of the Wigner parameters
\begin{gather}
e^{-i\beta}=|cz_2+d|^{2i},\qquad e^{-i\alpha}=\frac{\bar{c}\bar{z}_2+\bar{d}}{cz_2+d}.\label{wignerparametersforpairwisecelest}
\end{gather}
Returning to our expression for the Lorentz transformation of the pairwise celestial state (\ref{pairwisecelestwithoutexplicitparameters}) and plugging in the explicit form (\ref{wignerparametersforpairwisecelest}) for the Wigner parameters, we find that the pairwise celestial state transforms as
\begin{align*}
U[\Lambda]\ket{\set{z_1,z_2;\sigma_{12},\triangle_{12}}}
=\bigg(\frac{\bar{c}\bar{z}_2+\bar{d}}{cz_2+d}\bigg)^{\sigma_{12}}|cz_2+d|^{2i\triangle_{12}}\ket{\set{\Lambda z_1,\Lambda z_2;\sigma_{12},\triangle_{12}}}.\numberthis\label{assymetric}
\end{align*}
Although (\ref{assymetric}) does indeed form a representation of the Lorentz group, it exhibits an asymmetry between $z_1$ and $z_2$ that arose from our choice that $(0,\infty)\rightarrow (z_1,z_2)$ under the standard boost. To obtain a more symmetric representation we form two copies of the pairwise representation, one in which $(0,\infty)$ is mapped to $(z_1,z_2)$ under the standard boost, and the other where it is mapped to $(z_2,z_1)$. Taking the tensor product of these copies then gives the transformation law
\begin{align*}
U[\Lambda]\ket{\set{z_1,z_2;\sigma_{12},\triangle_{12}}}
=\bigg[\prod_{i=1,2}\bigg(\sqrt{\frac{cz_i+d}{\bar{c}\bar{z}_i+\bar{d}}}\bigg)^{\eta_1\eta_2\sigma_{12}}|cz_i+d|^{i\eta_1\eta_2\triangle_{12}}\bigg]\ket{\set{\Lambda z_1,\Lambda z_2;\sigma_{12},\triangle_{12}}}\numberthis\label{distributedtransformation}
\end{align*}
This is justified as the result (\ref{distributedtransformation}) again forms a unitary representation of the Lorentz group. We have also added sign factors $\eta_i$ to the exponents to distinguish in/out states. Let us comment on the apparent ``decoupling" exhibited in (\ref{wignerparametersforpairwisecelest}). Here, by apparent decoupling we refer to the fact that the Wigner parameters $\alpha,\beta$ do not take the naively expected form of a function of both directions e.g. $\alpha(z_1,z_2),\beta(z_1,z_2)$. We believe this feature is specific to the massless limit.  This is much like the situation in Section \ref{sectboostcharge} where the boost charge for a pair of massive charges (\ref{asymboost}) is a function of both momenta, however upon taking the massless limit, we found that each boost charge ``decoupled" into a form that only contained the direction of one of the particles.  \par
We now return to our original state (\ref{celesttriple}) and introduce the shorthand notation for our representation
\begin{align*}
&\ket{z_1,\triangle_1,\sigma_1;z_2,\triangle_2,\sigma_2;\triangle_{12},\sigma_{12}}\\
&\defined \ket{z_1;\triangle_1,\sigma_1}\otimes\ket{z_2;\triangle_2,\sigma_2}\otimes\ket{\set{z_1,z_2;\triangle_{12},\sigma_{12}}}\numberthis\label{triplestate}\\
&=\ket{z_1;\triangle_1,\sigma_1}\otimes\ket{z_2;\triangle_2,\sigma_2}\otimes\ket{z_1;\frac{\triangle_{12}}{2},\frac{\sigma_{12}}{2}}\otimes\ket{z_2;\frac{\triangle_{12}}{2},\frac{\sigma_{12}}{2}}\label{quadruple},\numberthis
\end{align*}
where in the last line (\ref{quadruple}) we have used that the transformation properties of the state (\ref{distributedtransformation}) are the same as that of a tensor product of two celestial representations with conformal dimensions $\frac{\triangle_{12}}{2}$ and helicities $\frac{\sigma_{12}}{2}$. Combining the transformation properties of all of the states involved, the state (\ref{triplestate}) transforms as
\begin{align*}
U[\Lambda]&\ket{z_1,\triangle_1,\sigma_1;z_2,\triangle_2,\sigma_2;\triangle_{12},\sigma_{12}}\\
&=\bigg[\,\prod_{i=1,2}\bigg(\sqrt{\frac{cz_i+d}{\bar{c}\bar{z}_i+\bar{d}}}\,\,\bigg)^{(2\eta_i\sigma_i-\eta_1\eta_2\sigma_{12})}|cz_i+d|^{i(\eta_{1}\eta_2\triangle_{12}-2(1+i\eta_i\lambda_i))}\bigg]\\
&\hspace{3.7cm} \times \ket{\Lambda z_1,\triangle_1,\sigma_1;\Lambda z_2,\triangle_2,\sigma_2;\triangle_{12},\sigma_{12}}.\label{finaltransformofpairwisecelest}\numberthis
\end{align*}
In summary, we have found that when Zwanziger's analysis is applied to massless celestial sphere representations, the net effect is very simply to shift the spin and conformal dimensions of each particle $i$ by a sum over pairwise terms
\begin{align}
(\triangle_j,\sigma_j)&\rightarrow \bigg(\triangle_j-i\sum_{k\neq j}\eta_k\frac{\triangle_{jk}}{2}\,\,,\,\,\sigma_k-\sum_{k\neq j}\eta_k\frac{\sigma_{kj}}{2}\bigg)\label{shiftofdimensions1}\\
&\defined(\tilde{\triangle}_j,\tilde{\sigma}_j),\label{shiftofdimensions2}\\
&\defined(1+i\tilde{\lambda}_j,\tilde{\sigma}_j)
\end{align}
where we have defined the shifted conformal dimension $\tilde{\triangle}_j$ and helicity $\tilde{\sigma}$ in the second line, and we have defined the shifted $\tilde{\lambda}$ parameter in the last line.  The pairwise nature of the representation is contained in the fact that the pairwise conformal dimension and helicities $\triangle_{ij},\,\,\sigma_{ij}$ involve parameters specific to the two individual particles.  We can already gain further insight into the $\triangle_{ij}$'s by noting that they must be Lorentz invariants of the two particles parameters, otherwise (\ref{finaltransformofpairwisecelest}) would no longer form a representation. In the massless celestial basis, where energy $E$ is no longer a variable, there are no such two particle Lorentz invariants (e.g. we cannot use  $p_1\.p_2$), hence the shift parameters must be pure numbers. In later sections we will find that these shift parameters in massless QED are just functions of the electric and magnetic charges of the particles.\par
 Despite the apparent simplicity of the prescription (\ref{shiftofdimensions1}), we will demonstrate in Section \ref{sectcompare} that this pairwise shift is precisely the effect that soft photon loop corrections and the FK dressing for massless charged particles in the celestial sphere representation were found to have in \cite{Arkani-from uv to IR}. 
\subsection{Celestial amplitudes for asymptotically interacting representations}\label{sectamplitudes}
In this section we construct celestial scattering amplitudes that are consistent with our modified transformation law (\ref{finaltransformofpairwisecelest}). The $SL(2,\mathbb{C})$ covariance of the states (\ref{finaltransformofpairwisecelest}) translates into the covariance of the amplitude in the following way\footnote{The set notation here is used to indicate the set of all particle labels and should not be confused with the set notation used to indicate a pairwise state.}
\begin{align*}
\A^{\text{C.S.}}\big(\set{z;\triangle,\sigma;\triangle_{ij},\sigma_{ij}}\big)&=\bigg(\prod_j|cz_j+d|^{-2-2i\tilde{\lambda}_j}\bigg(\frac{cz_j+d}{\bar{c}\bar{z}_j+\bar{d}}\bigg)^{\eta_j\tilde{\sigma}_j}\bigg)\\
&\times \A^{\text{C.S.}}\big(\set{\Lambda z;\triangle,\sigma;\triangle_{ij},\sigma_{ij}}\big),\numberthis\label{covofinteract}
\end{align*} 
 which differs from the usual $SL(2,\mathbb{C})$ covariance condition (\ref{amplitudecovariance}) by the replacement $(\triangle,\sigma)\rightarrow (\tilde{\triangle},\tilde{\sigma})$ in the exponents. We have also added particle labels $\triangle_{ij},\sigma_{ij}$ to the amplitudes. The reason for doing so here is different from the usual reason why one might add, for example the electric and magnetic charges $e,g$, as particle labels. In the usual approach, the electron charge $e$ serves as a free parameter in the amplitude, and it does not effect the Lorentz transformation properties of the states nor the Lorentz covariance of the amplitudes. However here we now add the labels $\triangle_{ij},\sigma_{ij}$, which we will soon show are related to the electric and magnetic charges, to index which pairwise principal series transformation law we are referring to.  With hindsight it should have been expected that $\triangle_{ij},\sigma_{ij}$ or equivalently $e,g$ should index the asymptotically interacting representations of the Poincar\'e group as these parameters distinguish an interacting state from a non-interacting state.\par
 It is straightforward to construct amplitudes consistent with the modified $SL(2,\mathbb{C})$ covariance property (\ref{covofinteract}). Let $\A_0^{\text{C.S.}}$ denote a celestial amplitude consistent with the unmodified covariance property (\ref{amplitudecovariance}) where there is no shift in the helicity or conformal dimensions. Let $\A^{\text{C.S.}}$ then denote the full amplitude which is consistent with the modified transformation law (\ref{covofinteract}). We will construct the $\A^{\text{C.S.}}$ from the $\A^{\text{C.S.}}_0$. There are three ways in which we can do so. The first method is straightforward; just shift the conformal dimension and helicities of $\A_0^{\text{C.S.}}$,
 \begin{equation}
 \A^{\text{C.S.}}(\set{z;\triangle,\sigma;\triangle_{ij},\sigma_{ij}})=A_0^{\text{C.S.}}(\set{z,\tilde{\triangle},\tilde{\sigma}}),\label{shiftamplitude}
 \end{equation}
 as $\A_0^{\text{C.S.}}$ satisfies the unshifted covariance condition (\ref{amplitudecovariance}), simply shifting the conformal dimensions and helicities in it's argument will result in an amplitude consistent with the modified transformation law (\ref{covofinteract}). In order to explain the second approach, first note the following covariance property
 \begin{equation}
 \Lambda z_i-\Lambda z_j=\frac{1}{(cz_i+d)(cz_j+d)}(z_i-z_j),\label{buildingblock}
 \end{equation}
 which the reader can readily verify upon making use of the $ad-bc=1$ condition. We therefore see that $z_{ij}\defined z_i-z_j$ is the building block for $SL(2,\mathbb{C})$ covariant amplitudes. Using this, we can construct an amplitude $\A^{\text{C.S.}}$ from an unmodified amplitude $\A^{\text{C.S.}}_0$ as follows
\begin{equation}
\A^{\text{C.S.}}(\set{z;\triangle,\sigma;\triangle_{ij},\sigma_{ij}})=\bigg(\prod_{i<j}\sqrt{\frac{z_{ij}}{\bar{z}_{ij}}}^{\,\,\eta_i\eta_j\sigma_{ij}}|z_{ij}|^{i\eta_i\eta_j\triangle_{ij}}\bigg)\A_0^{\text{C.S.}}(\set{z;\triangle,\sigma}).\label{modceles}
\end{equation}
Using the relation (\ref{buildingblock}) one can verify that (\ref{modceles}) satisfies the modified covariance property (\ref{covofinteract}). In this case, all of the effects of the asymptotic interactions are contained in the bracketed term in (\ref{modceles}). The factorization in (\ref{modceles}) is akin to that seen in the hard/soft factorization theorems seen in QED and gravity \cite{Yennie,Weinberg:1965nx}. In this case the factorization of the amplitude was possible due to the factorized nature of the Poincar\'e transformation properties.  A third option for constructing amplitudes, is to use a combination of these two approaches where we distribute the homogeneous scaling between the two: for example in (\ref{shiftamplitude}) we shift the helicities and conformal dimensions only by half of the shift indicated in (\ref{shiftamplitude}), and then multiply by the bracketed term in (\ref{modceles}) with the exponents then multiplied by a half. The covariance properties of the two terms then combine to form an amplitude with the correct total homogeneity. \par 
All three of these approaches result in different amplitudes, and one requires more information and context, such as a symmetry constraint or a requirement of IR finiteness, in order to determine which prescription is the relevant one for the model being studied.
\section{Comparison to IR divergences in massless QED}\label{sectcompare}
 In \cite{Arkani-from uv to IR} the effects of infrared divergences on celestial amplitudes in massless QED were studied. In this section we first briefly summarize some of the relevant key results from this work, and then demonstrate that the pairwise little group construction is consistent with, and provides an alternative rationale for these findings. We then compare these two approaches for obtaining IR finite celestial amplitudes. \par 
In \cite{Arkani-from uv to IR} the abelian exponentiation theorem \cite{Yennie} of momentum space amplitudes was Mellin transformed to a corresponding statement for celestial amplitudes. The celestial states were then dressed with an appropriate choice of Faddeev-Kulish dressing \cite{Kulish:1970ut}. It was concluded that when these two effects are taken into account, the IR finite dressed celestial amplitude is simply equal to the hard part of the celestial amplitude (to be defined below) with the conformal dimensions of the charged particles shifted in the following manner 
\begin{align}
\A^{\text{C.S.}}_{\text{dressed}}\big(\set{z_i;\triangle_i,\sigma_i}\big)&=\A_{\text{hard}}^{\text{C.S.}}\big(\set{z_i;\triangle_i+\eta_i\alpha Q_i^2,\sigma_i}\big)\label{monicashift}\\
\alpha&=\frac{e^2}{4\pi^2}\ln\Lambda_{IR}.
\end{align}
Here $\A^{\text{C.S.}}_{\text{dressed}}$ denotes the infrared finite celestial sphere (C.S.) amplitude where an appropriate FK dressing has been applied to the states.  $Q_i$ refers to the integer charges of the particles, $\eta=+/-$ for in/out states respectively, and $\Lambda_{IR}$ is the infrared cutoff. The prefactor $\frac{e^2}{4\pi^2}$ is the cusp anomalous dimension. Here, $\A_{\text{hard}}^{\text{C.S.}}(\set{z;\triangle,\sigma})$ is defined using the momentum space abelian exponentiation theorem, which states that the momentum space amplitude factorizes as
\begin{equation}
\A\Big(\set{p,\sigma}\Big)=e^{-\alpha\sum_{i<j}Q_iQ_j\ln|\frac{1}{2}p_i\. p_j|}\A_{\text{hard}}\Big(\set{p,\sigma}\Big).
\end{equation} 
We then define $\A^{\text{C.S.}}_{\text{hard}}(\set{z,\triangle,\sigma})$ as the Mellin transform of $\A_{\text{hard}}(\set{p,\sigma})$.\par  
Let us now compare (\ref{monicashift}) to our results of the previous sections. In Section \ref{sectamplitudes} we demonstrated how one can construct pairwise covariant celestial amplitudes $\A^{\text{C.S.}}$ from amplitudes $\A_0^{\text{C.S.}}$ which transform with the usual free particle covariance properties. Here we will take $\A_0^{\text{C.S.}}(\set{z,\triangle,\sigma})=\A_{\text{hard}}^{\text{C.S.}}(\set{z,\triangle,\sigma})$ which does indeed exhibit the free particle covariance properties. To construct the pairwise covariant amplitude we will use the method of shifting the conformal dimensions by a pairwise amount (\ref{shiftofdimensions1}), (\ref{shiftamplitude}). We set the pairwise helicity to zero $\sigma_{ij}=0$ as we are not considering the case of magnetic charges. We then make the following identification for the pairwise conformal dimension
\begin{equation}
\triangle_{ij}=-2i\alpha Q_iQ_j.\label{identificationofgamma}
\end{equation}  
Using this identification in (\ref{shiftofdimensions1}), and then making use of charge conservation, our shift simplifies to
\begin{align}
\tilde{\triangle}_i&=\triangle_i-\alpha Q_i\sum_{j\neq i}\eta_j Q_j\\
&=\triangle_i+\eta_i\alpha Q_i^2\label{pairwisesimplify}
\end{align}
Plugging this into our (\ref{shiftamplitude}) we obtain
 \begin{equation}
 \A^{\text{C.S.}}\big(\set{z_i;\triangle_i,\sigma_i;\triangle_{ij},\sigma_{ij}}\big)=A_0^{\text{C.S.}}\big(\set{z_i,\triangle_i+\eta_i\alpha Q_i^2,\sigma_i}\big).\label{final}
 \end{equation}
 We then identify the pairwise covariant amplitude $\A^{\text{C.S.}}$ with the IR finite dressed amplitude $\A^{\text{C.S.}}_{\text{dressed}}$ and upon doing so reproduce the result (\ref{monicashift}). In conclusion, we have found that when Zwanziger's construction is applied to massless celestial states, the analysis indicates that the helicity and conformal dimensions should shift by pairwise amounts. For massless QED without monopoles we make the identifications $\sigma_{ij}=0$ and (\ref{identificationofgamma}). Comparing (\ref{monicashift}) and (\ref{final}) we see that such a prescription correctly reproduces the infrared dynamics, namely the abelian exponentiation theorem and the FK dressing for celestial amplitudes in abelian gauge theories of massless charged particles.\par
 Let us emphasize some key differences between our approach and the methods used in \cite{Arkani-from uv to IR} for deriving (\ref{monicashift}). Firstly, our approach is almost entirely group theoretical.  We made no reference to Lagrangians, fields, creation and annihilation operators, or loop diagrams. In contrast, the method used in \cite{Arkani-from uv to IR} makes use of the abelian exponentiation theorem, which is arrived at by summing the $(e^2\ln\Lambda_{IR})^n$ contributions to the perturbation series which arise from soft virtual photon exchanges between charged particles. The shortcoming of our approach is that we are unable to predict the value of $\alpha$. \par 
 In light of the simplicity of the result (\ref{monicashift}), it appears in this case that the traditional Feynman perturbation series obscures the underlying covariance of the asymptotic states. To explain further on this point, let us use the identification (\ref{identificationofgamma}) in our transformation law (\ref{distributedtransformation}) for a pairwise state of two massless electrically charged particles
  \begin{equation}
  U[\Lambda]\ket{\set{z_1,z_2;Q_1,Q_2}}
  =\bigg[\prod_{i=1,2}|cz_i+d|^{2\eta_1\eta_2\,\alpha\, Q_1Q_2}\bigg]\ket{\set{\Lambda z_1,\Lambda z_2;Q_1,Q_2}}.
  \end{equation}
  The factors of the charge in the exponent highlight the non-perturbative, all orders in the coupling, nature of the result. This indicates a potential shortcoming of the traditional perturbation series.  If the $\ket{p,\sigma}$ states exhibit a similarly simple pairwise covariance where the coupling features in the exponent, then demanding that one works order by order in the coupling may overcomplicate the calculation of amplitudes. One of our motivations for seeking the correct asymptotic representations is that once the transformation properties are known, one should be able to develop a more efficient approach to calculating the amplitudes.  Indeed, one of the minimal inputs of the modern day ``on-shell methods" (see \cite{Elvang:2015rqa} for a review) approach to calculating scattering amplitudes is the transformation properties of the states. 
 \par
 Another feature of the Zwanziger approach is that we are able to discuss electric and magnetic charges simultaneously, in a rather straightforward manner. Indeed, the result (\ref{shiftofdimensions1}) suggests that the net effect of asymptotic electric-magnetic interactions $e_ig_j$ for massless dyons on the celestial is simply to shift the helicity of the two celestial states.\par
 Our approach may also aid the search for the generalization to the case of massive charged particles. The Zwanziger analysis suggests that one should search for a pairwise representation in the massive case \cite{L.Lippstreu}.  
 \par
 There is however an issue with the result (\ref{monicashift}). The shift in the conformal dimensions indicated in (\ref{monicashift}) knocks the states off of the principle series and hence no unitary inner product exists for these states \cite{Gelfand}. Our analysis only states that Zwanziger's little group implies a shift in the conformal dimensions, and does not determine whether the shift is complex or real. Unitarity then would imply that we should pick a complex shift for the conformal dimensions $\triangle_{ij}$. By matching to the result in \cite{Arkani-from uv to IR} we chose a real shift, in which case the representation is no longer unitary. We speculate that this problem has arisen because the traditional Feynman perturbation series assumes the incorrect transformation properties of the asymptotic states\footnote{An alternative resolution to this problem may be to use the results of \cite{c14Donnay:2020guq}, where it was shown that conformal primaries with general conformal dimensions can be expressed in terms of contour integrals over the $\triangle=1+i\mathbb{R}$ principal series.}. 

\section{Pairwise four-momentum representations}\label{sect,maingravity}
In the previous sections we have restricted our discussion to amplitudes where the matter particles in the asymptotic states are in the celestial sphere basis.  In Section \ref{sect,generalfourmomentum} we describe the construction when the matter particles are written in the four-momentum basis $\ket{p,\sigma}$. As an application of the formalism we will demonstrate in Section \ref{sect,gravity} that the gravitational and abelian exponentiation theorems exhibit the proposed pairwise covariance.  We have delayed the discussion of soft graviton exchanges until this section as these effects are much simpler when the asymptotic states are described in the four momentum basis as opposed to the celestial sphere basis, for reasons that should become apparent in Section \ref{sect,gravity}.  In \cite{c28Nande:2017dba} and \cite{Himwich:2020rro} it was demonstrated respectively that the exponential terms in the abelian and gravitational exponentiation theorems can be rewritten as correlators of vertex operators of the respective Goldstone bosons of the spontaneously broken asymptotic symmetries. In Section \ref{sect,vertex} we comment on the relation to the pairwise representations discussed herein. In particular we demonstrate that the insertion of vertex operators has the same effect on the amplitude as adding additional pairwise principal series representations to the asymptotic states.
\subsection{General massless pairwise representations in the four-momentum basis}\label{sect,generalfourmomentum}
Performing an inverse Mellin transform\footnote{To go from a celestial state to a four momentum state we perform an inverse Mellin transform
 \begin{equation}
 \ket{p,\sigma}=\int_{-\infty}^{\infty}\frac{\DD\lambda}{2\pi}\,\, \bigg(\frac{E}{1+|z|^2}\bigg)^{-i\eta\lambda-1} \ket{z,\triangle=1+i\lambda,\sigma}.
 \end{equation}} on the $\triangle_1$ and $\triangle_2$ variables of our result (\ref{quadruple}), we have that the net modification suggested by Zwanziger's pairwise little group is to add to each pair of massless momenta, two additional celestial sphere representations,
\begin{gather*}
\ket{p_1,\sigma_1}\otimes\ket{p_2,\sigma_2}\rightarrow \\
\ket{p_1,\sigma_1}\otimes\ket{p_2,\sigma_2}\otimes\ket{z_1;\triangle_{12},\sigma_{12}}\otimes\ket{z_2;\triangle_{12},\sigma_{12}}\numberthis\label{modstat}
\end{gather*}
where the pairwise nature of the representation is contained in the fact that the pairwise conformal dimension $\triangle_{12}$ and the pairwise helicity $\sigma_{12}$ involves parameters that are intrinsic to the pair. The additional states in (\ref{modstat}) transform as
\begin{align}
U[\Lambda]\ket{z_1;\triangle_{12},\sigma_{12}}&=|cz_1+d|^{i\eta_1\eta_2\triangle_{12}}\sqrt{\frac{cz_1+d}{\bar{c}\bar{z}_1+\bar{d}}}^{\,\,\eta_1\eta_2\sigma_{12}}\ket{\Lambda z_1;\triangle_{12},\sigma_{12}}\label{princ}.\\
\hat{P}^{\mu}\ket{z_1;\triangle_{12},\sigma_{12}}&=0\label{zeroenergy}
\end{align}
We motivate our choice (\ref{zeroenergy}) of declaring that the additional states are annihilated by the momentum operator both because we expect that these additional states arise due to zero energy graviton/photon exchanges, as well as our expectation that the asymptotic states in (\ref{modstat}) should have momentum eigenvalues degenerate with those of the free states.
Notice that (\ref{princ}) forms a representation of the Poincar\'e group so long as the pairwise conformal dimension $\triangle_{12}$ and helicity $\sigma_{12}$ are Lorentz invariants. This indicates a new possibility that opens up in the four momentum basis that was not present in the celestial sphere basis. In the four momentum basis we have the two-particle Lorentz invariant, Mandelstam $s_{ij}\defined p_i\. p_j$. Hence the pairwise conformal dimension and helicity can be a function of Mandelstam $s_{ij}$. This will be important when discussing gravity. In the celestial sphere basis it was not possible to form a two particle Lorentz invariant from the two stereographic directions of the particle pair. \par
Let us now construct amplitudes consistent with the transformation properties of the modified asymptotic states (\ref{modstat}). Let $\A_0$ denote the amplitude consistent with the transformation properties of the asymptotic states without the addition of the pairwise principal series states (first line of (\ref{modstat})), and let $\A$ denote the amplitude consistent with the transformation properties where the pairwise principal series representations have been added (second line of (\ref{modstat})). Then we can construct an $\A$ from an $\A_0$ simply by
\begin{equation}
\A\Big(\set{p,\sigma;\triangle_{ij},\sigma_{ij}}\Big)=\Bigg[\,\prod_{i,j}|z_{ij}|^{i\eta_i\eta_j\triangle_{ij}}\sqrt{\frac{z_{ij}}{\bar{z}_{ij}}}^{\,\,\eta_i\eta_j\sigma_{ij}}\Bigg]\A_0\Big(\set{p,\sigma}\Big).\label{constructamplitude}
\end{equation}
One can verify that $\A$ exhibits the correct Lorentz covariance properties by using the covariance property (\ref{buildingblock}) of the $z_{ij}$'s. The factorized nature of the RHS of (\ref{constructamplitude}) is akin to that seen in the hard/soft factorization theorems in abelian gauge theories and gravity \cite{Yennie,Weinberg:1965nx}, and in this case the factorization of the RHS was possible due to the factorized nature of the transform law for the states in (\ref{modstat}).\par
There is another possible contribution to $\A$ that should be mentioned. Now that we are working in the four momentum basis, we have access to the energy variable $E$, which can be combined with the stereographic direction to form the covariant quantity $\omega\defined \frac{E}{1+|z|^2}$. Under a Lorentz transformation $\Lambda={\scriptsize\begin{pmatrix}
a & b \\ 
c & d
\end{pmatrix}}$ this transforms as
\begin{equation}
\frac{(\Lambda E)}{1+|\Lambda z|^2}=|cz+d|^2\bigg(\frac{E}{1+|z|^2}\bigg)
\end{equation}
where $\Lambda E$ refers to the Lorentz transformed energy. Hence another covariant amplitude one could form is
\begin{equation}
\A\Big(\set{p,\sigma;\triangle_{ij},\sigma_{ij}}\Big)=\Bigg[\,\prod_{i,j}\omega_i^{-i\eta_i\eta_j\triangle_{ij}}\sqrt{\frac{z_{ij}}{\bar{z}_{ij}}}^{\,\,\eta_i\eta_j\sigma_{ij}}\Bigg]\A_0\Big(\set{p,\sigma}\Big).
\end{equation}
One could also take weighted products of the $\omega_i\omega_j$ and $|z_{ij}|$ terms to form covariant amplitudes. Which specific product one should take will be model dependent and requires additional informations such as a Lagrangian, an additional symmetry constraint, or a set of OPE's.\par
In order to establish a result to compare to in the next section, let us set $\sigma_{ij}=0$ and $\triangle_{ij}=i\gamma s_{ij}$ in (\ref{constructamplitude})
\begin{equation}
\A\Big(\set{p,\sigma}\Big)=\Big[\,\prod_{i,j}|z_{ij}|^{-\gamma\eta_i\eta_j s_{ij}}\Big]\A_0\Big(\set{p,\sigma}\Big),\label{graivty1}
\end{equation}
where $\gamma$ is an as of yet unspecified pure number which is unaffected by Lorentz transformations. Here we do not consider the option of using the $\omega_i$'s to form a covariant amplitude as in this case momentum conservation implies that $\sum_{j}\eta_j\triangle_{ij}=0$, hence $\prod_{i,j}\omega_i^{-i\eta_i\eta_j\triangle_{ij}}=1$, whereas no such simplification occurs for the $\prod_{i,j}|z_{ij}|^{\eta_i\eta_j\triangle_{ij}}$ terms.
\subsection{Gravitational and abelian exponentiation theorems}\label{sect,gravity}
The simplest example of a pairwise representation is one in which the pairwise conformal dimension is proportional to a product of two Lorentz invariant quantum numbers intrinsic to each individual particle $\triangle_{ij}=\alpha e_i e_j$. In Section \ref{sectcompare} we showed that this representation was equivalent to a particular choice of Faddeev-Kulish dressing in abelian gauge theories. The second simplest example of a pairwise representation is one in which the pairwise conformal dimension is proportional to the Lorentz invariant formed from the pairs' momenta $\triangle_{ij}=\gamma p_i\.p_j$. In this section we show that this choice reproduces the effects of soft graviton exchanges. \par
The leading order infrared divergences that result from summing  the exchange of soft gravitons between external legs of Feynman diagrams, where all loop momenta are cutoff above $\Lambda_{IR}$, exponentiate and factorize off from the rest of the momentum space amplitude $\A_0$ in the following manner \cite{Weinberg:1965nx}\footnote{The discussion from (\ref{exp}) to (\ref{pdot}) is taken from \cite{Arkani-from uv to IR}.}
\begin{equation}
\A=e^B\A_0\label{exp}
\end{equation}
where $\A_0$ does not involve $\Lambda_{IR}$. In this case the exponent is \cite{Naculich:2011ry}
\begin{align}
B&=-\gamma\sum_{i,j}\eta_i\eta_j(p_i\. p_j)\ln\bigg(\frac{p_i\.p_j}{\mu^2}\bigg)\label{firstline}\\
&=-\gamma\sum_{i,j}\eta_i\eta_j s_{ij}\ln|z_{ij}|^2\label{secondlines}
\end{align}
with 
\begin{equation}
\gamma=\frac{G}{\pi}\ln\Lambda_{IR}.\label{gamma}
\end{equation}
and $\mu^2$ is an arbitrary mass scale and $s_{ij}\defined p_i\. p_j$. To get the second line (\ref{secondlines}) we used 
\begin{equation}
p_i\.p_j=2\omega_i\omega_j|z_{ij}|^2.\label{pdot}
\end{equation}
and then used momentum conservation $\sum_i \eta_i p_i^{\mu}=0$. We can therefore rewrite the gravitational exponentiation theorem as
\begin{equation}
\A=\bigg(\prod_{i,j}|z_{ij}|^{-\gamma\eta_i\eta_j s_{ij}}\bigg)\A_0\label{gravity2}
\end{equation}
We see that (\ref{graivty1}) and (\ref{gravity2}) are the same, thus indicating that one can view the gravitational exponentiation theorem as arising due to an underlying \textit{pairwise} covariance of the asymptotic states. Four clarifying comments are in order:\par
(1) Let us first note the Lorentz covariance properties of each one of the factors in brackets in (\ref{gravity2})
\begin{equation}
|\Lambda z_i-\Lambda z_j|^{-\gamma\eta_i\eta_j s_{ij}}=|cz_i+d|^{\gamma\eta_i\eta_j s_{ij}}|cz_j+d|^{\gamma\eta_i\eta_j s_{ij}}|z_i-z_j|^{-\gamma\eta_i\eta_j s_{ij}}\label{covofKoba}
\end{equation}
where the $\gamma s_{ij}$ exponents on the $|cz_i+d|$ factors indicate the intrinsically pairwise nature of the transformation law. However let us note that on the support of momentum conservation, the net transformation law for the total amplitude is the same as that for the amplitude without the modified transformation law. For example, the covariance equation for the amplitude will contain a factor of
\begin{equation}
\prod_{i,j}|cz_i+d|^{\eta_i\eta_j\gamma s_{ij}}=\prod_i|cz_i+d|^{\gamma\eta_i p_i\.\sum_{j }\eta_jp_j}=1,
\end{equation} where we used momentum conservation $\sum_i\eta_i p_i^{\mu}=0$ in the last equality. One might then argue that it is not beneficial to point out the pairwise covariance of the states as the total amplitude does not exhibit the pairwise covariance. However this line of reasoning would also imply that it is not beneficial to point out that the momentum operator acts as $e^{i\hat{P}\.a}\ket{p^{\mu}}=e^{ip\.a}\ket{p^{\mu}}$ on individual states seeing that $e^{i\hat{P}\.a}$ acts as the identity operator on the amplitude on the support of momentum conservation. The benefit of pointing out the covariance of the states in both cases is that it indicates which wavefunctions/intertwiners and operators one should use in the field theory description. In the case of the momentum operator, the covariance of the individual states indicates that one should use the $e^{ip\.x}$ wavefunction/intertwiner in the field theory description of local interactions. In the case of a pairwise covariance law, such as (\ref{princ}), the pairwise covariance indicates that one should use vertex operators $e^{i\alpha \hat{C}(z,\bar{z})}$ in the field theory description. We explain further on this point in the next subsection. \par
(2)  The analogous argument for the massless abelian exponentiation theorem is ambiguous due to the presence of collinear divergences. In the case of gravity this is not an issue as the collinear divergences automatically cancel amongst one another \cite{Weinberg:1965nx}. The massless abelian exponentiation theorem, which sums the leading infrared divergences due to soft virtual photon exchanges between external legs of Feynman diagrams, is structurally the same as in (\ref{exp}) with the exponent taking the form \cite{Weinberg:1965nx}
\begin{align}
B&=-\frac{\alpha}{2}\sum_{i,j}\eta_i\eta_jQ_iQ_j\ln\bigg(\frac{p_i\.p_j}{\mu^2}\bigg)\label{abelian}\\
&=-\frac{\alpha}{2}\sum_{i,j}\eta_i\eta_jQ_iQ_j\bigg(\ln(|z_i-z_j|^2)+\ln\bigg(2\frac{\omega_i\omega_j}{\mu^2}\bigg)\bigg)\label{abelian2}\\
&=-\frac{\alpha}{2}\sum_{i,j}\eta_i\eta_jQ_iQ_j\ln(|z_i-z_j|^2)\label{abelian3}
\end{align}
where $\alpha=\frac{e^2}{4\pi^2}\ln\Lambda_{IR}$, $\mu$ is an arbitrary mass scale, $Q_i$ refers to the integer charge of the particles, we used (\ref{pdot}) in going to (\ref{abelian2}), and we used charge conservation $\sum_{i}\eta_iQ_i=0$ in going to (\ref{abelian3}). In \cite{Arkani-from uv to IR} the collinear divergences were dropped in a Lorentz invariant manner by discarding the $i=j$ terms in the sum (\ref{abelian}). However, if we drop the collinear divergences in a Lorentz covariant manner by discarding the $i=j$ terms in the sum (\ref{abelian3}), then the remaining terms in $e^B$ will take the form
\begin{equation}
\prod_{i<j}|z_{ij}|^{-\alpha\eta_i\eta_jQ_iQ_j}.\label{Coulombgas}
\end{equation}
The individual terms in the product (\ref{Coulombgas}) transform under Poincar\'e transformations in exactly the same manner as the additional pairwise celestial states in (\ref{modstat}) with pairwise conformal dimension $\triangle_{ij}=i\alpha Q_iQ_j$ and pairwise helicity $\sigma_{ij}=0$. Thus if one drops the collinear divergences in this manner then the massless abelian exponentiation theorem can be interpreted as arising from the pairwise covariance of the asymptotic states. Studying massive pairwise representations \cite{L.Lippstreu} should settle whether this viewpoint is justified.\par
(3) It is interesting to note that the exponents in representations of the Lorentz group of the form (\ref{princ}) are identified with the Casimirs of the Lorentz group \cite{Gelfand}. In the cases of gravity and electromagnetism we have identified the pairwise conformal dimensions with functions proportional to $G_N$ and $e^2$ respectively. Thus, in this approach Newton's constant and the particle's charge are identified with Casimirs of the Lorentz group. This would be a pleasing result, as Newton's constant and a particle's electric charge have the property that they are invariant under Lorentz transformations, which is precisely the property of a Casimir. However, due to the issues discussed at the end of this section, it is not immediately apparent whether this viewpoint is justified. \par
(4) There is an error in the philosophy of looking at known amplitudes to extract new transformation laws. First note that the Feynman rules are hardwired to produce amplitudes which exhibit the covariance properties of Wigner's one-particle irreducible representations. That is, amplitudes derived using the usual Feynman rules for QED or gravity will always transform as a tensor product of free particle representations, even if the late time interactions lead to the actual asymptotic states not transforming in this manner. This is the reason why the exponential factors in the gravitational and abelian exponentiation theorems are Lorentz invariant, as opposed to Lorentz covariant, e.g. (\ref{firstline}). It is a curious accident that momentum conservation allowed us to go from an expression (\ref{firstline}) where every term is Lorentz invariant, to (\ref{secondlines}) where each term in the sum exhibits a pairwise Lorentz covariance equivalent to that of a pair of pairwise celestial sphere representations, which only when summed together combine to give a Lorentz invariant quantity. One cannot expect this to be a generic feature; it is unreasonable to expect that one can use a set of Feynman rules that assume the incorrect Poincar\'e transformation properties to obtain an amplitude, and then use some conservation laws such as momentum conservation or charge conservation to rewrite the amplitude in a form that exhibits the correct covariance properties of the asymptotic states. The novel transformation properties of the asymptotic states arise in QED and gravity when one modifies the asymptotic states by inserting vertex operators, Faddeev-Kulish dressings or Wilson lines, thereby adding states with new transformation properties.  Thus a more appropriate way of determining the transformation properties of the asymptotic states is to analyze the covariance properties of amplitudes that contain these additional operator insertions. For example, consider adding coherent states to the in/out states of the amplitude in (\ref{exp}). These coherent states are designed to cancel the infrared divergence that occurs when $\Lambda_{IR}\rightarrow 0$, hence the modification to the amplitude will be to multiply by another factor of $e^B$ but with the logarithm in (\ref{gamma}) replaced by $\log\Lambda_{IR}\rightarrow\log\frac{\lambda}{\Lambda_{IR}}$, where $\lambda$ is an energy scale specific to the choice of dressing in the coherent state. Combining the two factors of $e^B$ we see that the net effect of adding this choice of coherent state would simply be to replace $\Lambda_{IR}\rightarrow \lambda$. Hence in this circumstance we arrive back at the same pairwise transformation properties for the amplitude. However, more generally other choices of dressing could give rise to additional Lorentz covariant factors.
\par
There are two issues with our construction that need to be resolved in future works:\par
(1) In order for (\ref{princ}) to form a representation of the Lorentz group we require $\triangle_{12}$ and $\sigma_{12}$ to be Lorentz invariant. However, here we are identifying $\triangle_{ij}=-2\gamma s_{ij}$ where $\gamma$ contains the logarithm of the IR cutoff $\log\Lambda_{IR}$. In most discussions of infrared divergences the Lorentz transformation properties of the cutoff $\Lambda_{IR}$ are glossed over.  However, as it is an energy scale we cannot expect it to be Lorentz invariant. We speculate that this contradiction has arisen because we started with a set of Feynman rules that assumed the incorrect transformation properties of the asymptotic states, and that this contradiction may be resolved if one starts with a set of Feynman rules that have the correct transformation properties built in.\par
(2) The pairwise representation with $\triangle_{ij}=i\gamma s_{ij}$ does not lie on the principal series hence no unitary inner product exists for these states \cite{Gelfand}. A potential resolution to this problem may be to use the results of \cite{c14Donnay:2020guq}, where it was shown that conformal primaries with general conformal dimensions can be expressed in terms of contour integrals over the principal series.
\subsection{Vertex operators}\label{sect,vertex}
In \cite{c28Nande:2017dba} and \cite{Himwich:2020rro} it was demonstrated that the soft factors in the abelian and gravitational exponentiation theorems can be rewritten as correlation functions of vertex operators of the Goldstone bosons of the respectively broken large gauge and supertranslations asymptotic symmetries. In this section we demonstrate the straightforward relation to the pairwise celestial representations. We demonstrate that the additional states in (\ref{modstat}) can be identified with the states produced by these vertex operators, as they imply the same covariance properties for the amplitude. This relation offers an alternate viewpoint on vertex operators, namely that their primary function is to implement the pairwise Poincar\'e covariance of the asymptotic states.  We discuss the case of gravity in this section with parallel remarks applying for the case of QED. \par  
 In \cite{Himwich:2020rro} it was demonstrated that the soft factor in the gravitational exponentiation theorem (\ref{exp}) can be re-expressed as a correlation function of vertex operators. They considered the following correlation function 
\begin{equation}
\braket{e^{i\eta_1\omega_1\hat{C}(z_1,\bar{z}_1)}...e^{i\eta_i\omega_i\hat{C}(z_i,\bar{z}_i)}}\label{correlator}
\end{equation}  
where $C(z,\bar{z})$ is the Goldstone boson of spontaneously broken supertranslations and $\omega_i\defined \frac{E_i}{1+|z_i|^2}$ is defined as before.  They determined that the only non-zero connected $n$-point function of $C(z,\bar{z})$ is the two-point function, which was determined to be 
\begin{equation}
\braket{\hat{C}(z,\bar{z})\hat{C}(w,\bar{w})}=-4\gamma |z-w|^2\ln|z-w|^2,
\end{equation}
in which case the  correlator (\ref{correlator}) evaluates to
\begin{align}
\braket{e^{i\eta_1\omega_1\hat{C}(z_1,\bar{z}_1)}...e^{i\eta_i\omega_i\hat{C}(z_i,\bar{z}_i)}}&=\exp\Bigg(-2\gamma\sum_{i\neq j}\eta_i\eta_j\omega_i\omega_j |z_{ij}|^2\ln|z_{ij}|^2 \Bigg)\label{cor1}\\
&=\prod_{i\neq j}|z_{ij}|^{-\eta_i\eta_j\gamma s_{ij}}\label{cor2}.
\end{align}
Observing that (\ref{cor2}) is the same as the term in brackets in (\ref{gravity2}), the authors of \cite{Himwich:2020rro} conclude that the soft factor in gravity can be re-expressed as a correlator of vertex operators of the Golstone boson of spontaneously broken supertranslations. The relation to our work is straightforward: each term in the product (\ref{cor2}) transforms under the Poincar\'e group as a pair of pairwise celestial representations with pairwise conformal dimension $\triangle_{ij}=i\gamma s_{ij}$. That is, the amplitude produced by the vertex operators (\ref{correlator}) exhibits the same pairwise covariance as adding pairwise celestial representations to the asymptotic states, as in (\ref{modstat}).   As we have indicated in Section \ref{sect,generalfourmomentum} there are multiple amplitudes consistent with adding additional pairwise celestial representations, and one requires more information, such as a Lagrangian or an additional symmetry constraint, to specify which precise amplitude is relevant. In the case of the vertex operators this ambiguity translates to fixing the $n$-point functions of the $C(z,\bar{z})$ operators.  Two comments are in order:\par
(1) This correspondence between our additional states in (\ref{modstat}) and the vertex operators of the $\hat{C}(z,\bar{z})$ Goldstone boson strengthens our choice in (\ref{zeroenergy}) that the additional pairwise states are annihilated by the momentum operator. Indeed, the $\hat{C}(z,\bar{z})$ field is composed of zero energy gravitons \cite{A2He:2014laa}, and thus these vertex operator insertions also do not impart momentum to the states.  \par
(2)  This correspondence between vertex operators and pairwise celestial representations emphasizes an alternate interpretation of vertex operators. Namely that they are a field theoretic means for implementing an underlying pairwise covariance of the asymptotic states.\par
To summarize this section: in order to encode the Poincar\'e transformation properties of the asymptotic states one appends to each pair of massless matter particles an additional pair of celestial states with pairwise conformal dimensions and helicities (\ref{modstat}). These additional states reflect the pairwise nature of the Poincar\'e transformation law. We demonstrated that if one makes this modification, then the two simplest choices for the pairwise conformal dimensions $\triangle_{ij}=i\gamma s_{ij}$ and $\triangle_{ij}=i\alpha e_ie_j$, reproduce the soft factors in the gravitational and, tentatively, depending on how one handles the collinear divergences, the abelian exponentiation theorems as tree-level amplitudes. An effective means for implementing such a pairwise covariance is by the use of vertex operators. Appending all of these additional states to the in/out states is cumbersome, thus a more pragmatic approach would be to recognize the pairwise nature of the in/out states, and then use vertex operators to implement this pairwise covariance. This observation provides an alternative explanation for why the authors of \cite{Himwich:2020rro,c28Nande:2017dba} were able to reproduce the soft factors in the abelian and gravitational exponentiation theorems as correlators of vertex operators. 

\section*{Acknowledgments}
I am especially grateful for many helpful conversations with Marcus Spradlin and Anastasia Volovich, as well as for providing valuable feedback on draft versions of the manuscript. I would also like to thank Chung-I Tan, Antal Jevicki, Jorge Mago and Hofie Hannesdottir for many useful discussions, and acknowledge Ana-Maria Raclariu and Shamik Banerjee for helpful correspondences. This work was supported in part by the US Department of Energy under
contract {DE}-{SC}0010010 Task A.

\providecommand{\href}[2]{#2}\begingroup\raggedright\endgroup

\end{document}